\documentclass[twocolumn]{pasj00}

\begin{document}
\SetRunningHead{IHARA et al.}{Searching for a companion of SNIa}
\Received{2006/07/19}
\Accepted{2007/04/23}

\title{Searching for a Companion Star of Tycho's Type Ia Supernova 
with Optical Spectroscopic Observations%
\thanks{Based on data collected at Subaru Telescope, which is operated by the National Astronomical Observatory of Japan.}}

\author{Yutaka \textsc{Ihara},\altaffilmark{1,2}
	Jin \textsc{Ozaki},\altaffilmark{2,3}
        Mamoru \textsc{Doi},\altaffilmark{1}
        Toshikazu \textsc{Shigeyama},\altaffilmark{3}
        \\
        Nobunari \textsc{Kashikawa},\altaffilmark{4}
        Yutaka \textsc{Komiyama},\altaffilmark{4}
        and 
        Takashi \textsc{Hattori},\altaffilmark{5}
        }
\altaffiltext{1}{Institute of Astronomy, University of Tokyo, 2-21-1 Osawa, Mitaka-shi, Tokyo 181-8588}
\altaffiltext{2}{Department of Astronomy, Graduate School of Science, University of Tokyo, Bunkyo-ku, Tokyo 113-0033}
\altaffiltext{3}{Research Center for the Early Universe, Graduate School of Science, 
\\ University of Tokyo, Bunkyo-ku, Tokyo 113-0033}
\altaffiltext{4}{National Astronomical Observatory of Japan, 2-21-1 Osawa, Mitaka-shi, Tokyo 181-8588}
\altaffiltext{5}{Subaru Telescope, National Astronomical Observatory of Japan, 
\\ 650 North A'ohoku Place, Hilo, HI 96720, USA} 
\email{iharayt@ioa.s.u-tokyo.ac.jp}


%

\KeyWords{supernova Ia: supernova remnants: Tycho} 

\maketitle

\begin{abstract}
We report our first results of photometric and 
spectroscopic observations for Tycho's supernova remnant (SNR Tycho) 
to search for the companion star of a type Ia supernova (SN Ia). 
From photometric observations using Suprime-Cam on the Subaru Telescope, 
we have picked up stars brighter than 22 mag (in $V$-band) for spectroscopy, 
which are located within a circular region with the radius of 30" around the center of SNR Tycho. 
If the ejecta of young supernova remnants, such as SNR Tycho, have a sufficient amount of Fe I, 
we should be able to detect absorption lines at 3720 \AA\, and 3860 \AA\, associated with transitions 
from the ground state of Fe I in the spectrum of the companion star.
To identify the companion star of a SN Ia using these characteristic absorption lines of Fe I, 
we made optical low-resolution spectroscopy of their targets using FOCAS on the Subaru Telescope. 
In our spectroscopic observations, 
we obtained spectra of 17 stars in the SNR Tycho region and compare them with template stellar spectra. 
We detect significant absorption lines from two stars at 3720 \AA. 
In addition, spectra of four stars also have possible absorption lines. 
Since widths of their absorption lines are broad, 
it is likely that the detected absorptions are due to Fe I in the expanding ejecta of SNR Tycho. 
However, none of stars exhibits a clear red wing in the observed profiles of the absorption, 
though a star in the background of the SNR should show it. 
Hence, we suggest another interpretation that the detected absorption lines might be caused by the peculiarity of stars.
Though Tycho(G) was raised as a candidate of the companion star of Tycho's supernova in a previous research, 
we do not obtain evidence supporting the previous research. 
On the contrary, a star named Tycho(E) has the absorption line at 3720 \AA\,  
and its projected position is close to the center of SNR Tycho.  
Based on our observations, 
Tycho(E) is a new candidate as the companion star of Tycho's supernova.

\end{abstract}

\section{Introduction}
Type Ia supernovae (SNe Ia) have been used as standard distance indicators and have become a powerful tool 
for studying the expansion of the Universe, since they exhibit similar maximum brightness and uniform light curves. 
They are used to constrain the cosmological parameters and 
provide the evidence for the existence of dark energy (Schmidt et al. 1998;Perlmutter et al. 1999; Knop et al. 2003; Astier et al. 2006; Riess et al. 2006).

Despite of their importance in the cosmological studies, the progenitor system of SNe Ia has not been fully understood yet and is still debated (Branch et al. 1995). 
There are two models for the progenitor system, which are the double degenerate model (DD) 
and the single degenerate model (SD). 
The DD scenario assumes that merging of double C+O white dwarfs with the combined mass exceeding the Chandrasekhar 
mass limit, which is about 1.4 $M_\odot$, results in SN Ia explosion (e.g., Iben \& Tutukov 1984; Webbink 1984).
However, the DD has been suggested to lead to an accretion-induced collapse rather than a SN Ia (e.g., Nomoto \& Iben 1985; 
Saio \& Nomoto 1985). 
On the other hand, in the SD scenario,
a white dwarf accretes the gas from the companion non-degenerate star in a binary system 
and increases its mass up to the Chandrasekhar limit to explode as a SN (e.g., Nomoto et al. 1997; Han \& Podsiadlowski 2004). 
Hachisu, Kato, \& Nomoto (1996) proposed a new plausible progenitor system based on the optically thick wind 
theory of mass-accreting white dwarfs. 
In this model, the accretion to the white dwarf at a rate exceeding the rate of hydrogen burning near the surface is inhibited by blowing a strong wind. 
In this way, the wind avoids the formation of a common envelope 
and the white dwarf can increase its mass up to the Chandrasekhar limit. 
They suggested two main evolutionary paths to SNe Ia, depending on the evolutionary stage of the companion star: a helium-rich supersoft X-ray source channel (Hachisu et al. 1999a) with the companion of a main sequence star 
and a wide symbiotic channel (Hachisu et al. 1999b) with a red giant. 

The above evolutionary scenarios for SNe Ia can be tested by observations. 
For example, a companion star should remain in the vicinity of the  
explosion site after the explosion of a SN Ia in the SD scenario.
If the companion star is identified by observations, the SD scenario will be supported. 

Ruiz-Lapuente et al. (2004) (hereafter Ruiz04) reported results of their survey of the central region of SNR Tycho to search for the companion star.
They measured velocities and distances of stars in the vicinity of the center of the SNR and 
found a G2IV star moving at a velocity more than three times greater than the velocity dispersion of disk stars in that region.  
As a result, they claimed that this star was the companion star. (They called this star "Tycho(G)".) 
This type of companion stars contradicts the evolutionary paths proposed by Hachisu et al. (1999a, b) 
and seems to suggest an alternative evolutionary path to a SN Ia.  
Fuhrmann (2005) has argued that Tycho(G) might be a thick-disc star which is coincidentally 
passing the vicinity of the supernova remnant, though the probability is  
extremely low.   
Therefore, the physical association of Tycho(G) with the SNR needs to be confirmed by other methods. 
To this end, we performed photometric and spectroscopic observations of stars including Tycho(G) 
with SUBARU telescope and present our results in this paper. 

Ozaki \& Shigeyama (2006) (hereafter OS06) proposed an alternative method to investigate the association of a star with the SN ejecta. 
When a star is in the vicinity of the center of the expanding ejecta, photons emitted from the star toward the observer are absorbed only by the ejecta 
moving toward the observer. Hence the absorption lines exhibit broad wings present only in the blue-shifted side, while a background star exhibits very broad wings in both of the blue and red sides 
of the absorption lines. 
The possibility that another irrelevant star happens to be inside the ejecta of SNR Tycho is only 0.1 \% as estimated from the density of stars there. 
Therefore a star inside SNR Tycho is very likely to be the companion star of Tycho's SN. 
In SNR Tycho, atoms are expected to be in the ground states. 
Thus useful absorption lines in the optical wavelength range can be formed by Fe I at 3720 \AA$\,$ and 3860 \AA. 

Tycho's supernova (SN 1572) is one of the type Ia supernovae in our galaxy.  
The distance to SNR Tycho is $\sim3$ kpc, derived from its X-ray spectrum (Fink et al. 1994). 
From the X-ray imaging observations with Chandra, the center of 
the remnant is determined to be ($\alpha$, $\delta$) = (\timeform{00h25m19.40s}, \timeform{64D08'13''.98}) (J2000.0) 
to minimize the ellipticity of the blast wave of the SNR (Warren et al. 2005). 
They also estimated the diameter of SNR Tycho as 8'. 
SNR Tycho is one of a few well-studied young remnants with known age and  
the remnant is so young that it could retain enough amount of Fe I 
to form absorption lines in the spectrum of the companion star.
 
In \S2, we report two kinds of observations, imaging and spectroscopy. 
In \S2.1, we describe results of imaging observations and explain how we determine the region to search 
the companion star. In \S2.2, we describe results of spectroscopic observations and data reduction. 
In \S3, we explain the method to estimate the amount of absorption lines of Fe I in terms of equivalent widths.
In \S4, we summarize the results of our observations. 
Finally, we discuss the detected absorption lines 
and the possibility of identification of the companion star of Tycho's SN in \S5.

\section{Observations}
\subsection{Photometric Observations}
\subsubsection{Observations}
Stars in the region of SNR Tycho was observed with Suprime-Cam (Miyazaki et al. 2002) at the prime focus of 8.2 m Subaru telescope 
on Nov. 4, 2002. Suprime-Cam is equipped with ten MIT/LL 2048 $\times$ 4096 CCDs 
arranged in a 5 $\times$ 2 pattern to provide 34' $\times$ 27' FOV with a pixel size on 
the sky of 0.201" $\times$ 0.201". We used $V$ and $R_\mathrm{c}$ filters of Johnson-Cousins systems and $i'$ filter of SDSS 
(Fukugita et al. 1996).
Three 10 sec and three 100 sec exposures were taken using each filter. 
We divided exposure into three and dithered to remove cosmic rays and to avoid bad pixels. 
The individual dithered images were coadded. The seeing size was about 0.8". 

\vspace{1em}

\subsubsection{Reduction and Photometry}
We analyze a 6' $\times$ 6' region around the center of SNR Tycho in images taken with Suprime-Cam. 
The size of the field is slightly larger than the field of view of FOCAS. 
The standard data reduction, including overscan subtraction and flatfielding, is performed 
with reduction software nekosoft for Suprime-Cam (Yagi et al. 2002). 
We smooth images with Gaussian kernel and search peaks to identify 
stellar positions. 
If the signal-to-noise ratio of a peak is higher than 5, then we identify the peak as an object. 
Photometry is then performed with a 2" $\phi$ aperture. 
Since we did not observe standard stars, the flux calibration of $V$ and $R_\mathrm{c}$-band is performed with the published photometry 
in SNR Tycho field (Ruiz04). For $i'$-band, we transform $i'$ mag into $I_\mathrm{c}$ mag of Johnson-Cousins systems 
by using the prescription in Fukugita et al. (1996). To calculate the zero point of $I_\mathrm{c}$ mag, 
we use $B$ and $V$ mag of Ruiz04 and the color of $B - V$ and $V - I_\mathrm{c}$ of Soderblom et al. (1998). 
To avoid saturations and to increase signal-to-noise ratio as much as possible, 
the total exposure times are chosen to be 30 sec for bright objects with the magnitudes of 16.0-18.5 mag ($V$), 
and 300 sec for faint objects with the magnitudes of 18.5-24.0 mag.
The limiting magnitudes (5 $\sigma$) are 24.0 mag ($V$), 23.5 mag ($R_\mathrm{c}$), and 23.5 mag ($I_\mathrm{c}$) for 300 sec. 
We make a star catalog from these images. 

\vspace{1em}

\subsubsection{Results of Photometry}
\begin{figure}
  \begin{center}
    \FigureFile(80mm,80mm){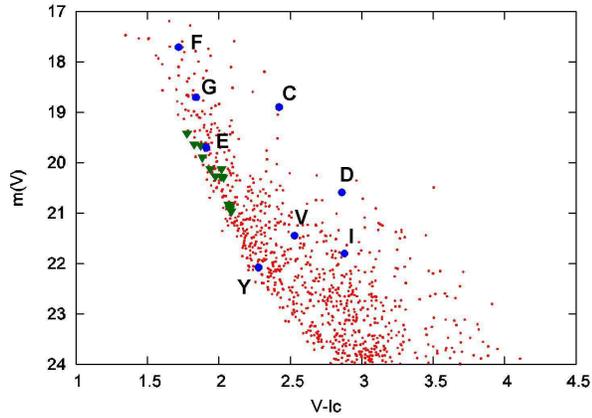}
  \end{center}
  \caption{Color-magnitude diagram in $V - I_\mathrm{c}$ versus $V$. m($V$) is the apparent magnitude in $V$-band. 
  Filled blue circles are stars in the candidate region. 
  Filled green triangles are observed stars as candidates of background stars of SNR Tycho. 
  We select blue stars as much as possible, because they are luminous.  
   }\label{fig:1}
\end{figure}

\begin{figure}
\begin{center}
    \FigureFile(80mm,80mm){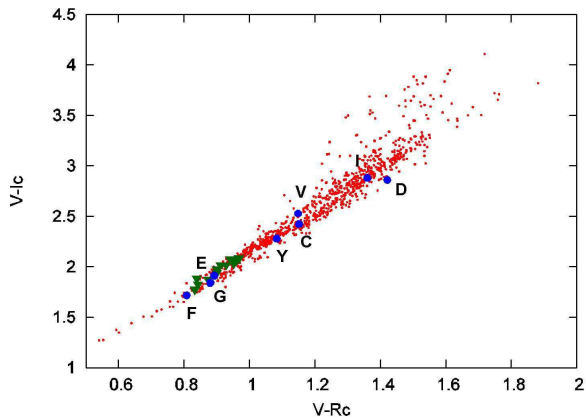}
  \end{center}
  \caption{Color-Color diagram in $V - R_\mathrm{c}$ versus $V - I_\mathrm{c}$.
  Filled blue circles are stars in the candidate region. 
  Filled green triangles are observed stars as candidates of background stars of SNR Tycho. 
  }\label{fig:1a}
\end{figure}

\begin{figure}
  \begin{center}
    \FigureFile(80mm,80mm){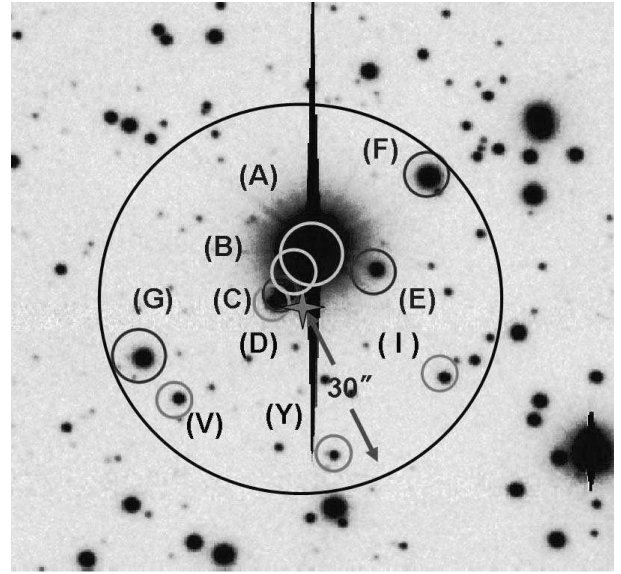}
  \end{center}
  \caption{A $V$-band image around the center of SNR Tycho taken with Suprime-Cam for 300 sec exposure. 
  The 30" radius circle is the candidate region. 
  The stars designated with capital alphabets are candidate stars of a companion star of Tycho's SN. 
  The center of SNR Tycho is marked by a gray star symbol. 
  }\label{fig:2}
\end{figure}

\begin{figure}
  \begin{center}
    \FigureFile(80mm,80mm){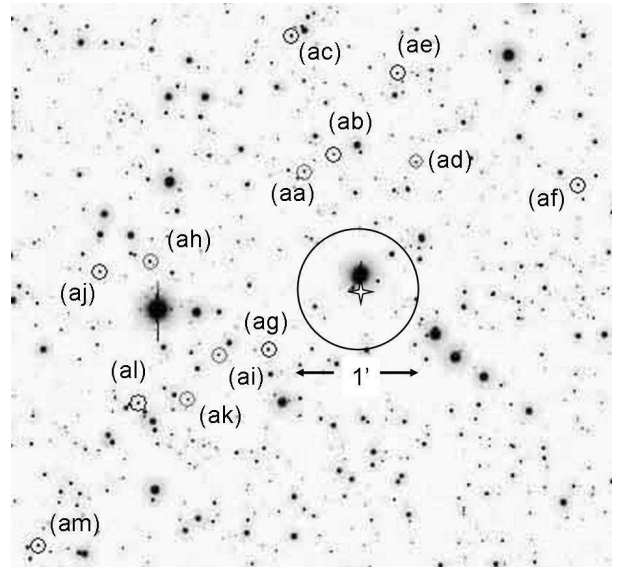}
  \end{center}
  \caption{A $V$-band image around the center of SNR Tycho with Suprime-Cam for 300 sec exposure. 
  The area of the image is almost the same as the field of view of spectroscopic observations with FOCAS.
  The 1' diameter circle shows the candidate region for the companion star. 
  The stars designated with lower case alphabets are spectroscopically observed 
  to detect Fe I in the SNR. 
  The center of SNR Tycho is marked by a star symbol. 
  }\label{fig:3}
\end{figure}

The final catalog contains 904 stars with photometry in $V$,$R_\mathrm{c}$, and $I_\mathrm{c}$ bands. 
This catalog is limited by $V$-magnitudes.  
A color-apparent magnitude diagram of $V - I_\mathrm{c}$ versus $V$ is shown in Figure \ref{fig:1} and 
a color-color diagram of $V - R_\mathrm{c}$ versus $V - I_\mathrm{c}$ is shown in Figure \ref{fig:1a}. 
We show a $V$-band image taken with Suprime-Cam around the center of SNR Tycho in Figure \ref{fig:2} 
and a $V$-band image which corresponds to the field of view of our spectroscopic observations in Figure \ref{fig:3}. 
These observations are described in detail in \S 2.2. 

\vspace{1em}


\subsubsection{Candidate region for the companion star}
We select candidates for the companion star of Tycho's SN using the projected distances 
from the center of SNR Tycho. 
We estimate the velocity of the companion star from the equation of motion of two bodies. 
In the SD model, 
there should have been a stellar binary system in which a white dwarf and the companion star revolve around their center of mass before SN Ia explosion. 
We calculate the orbital velocity of the companion star using the final mass and the period presented in Hachisu \& Kato (2001). 
We obtain the velocity of $\sim$200 km/s or less. 
After the explosion, only the companion star remains and is supposed to move with the same velocity. 
Four hundred and thirty years have passed since the explosion of Tycho SN1572. 
The distance from the Earth to SNR Tycho is $\sim$3 kpc (\S1). 
As a result, the companion star may have moved at most by $\sim$10" from the center of SNR Tycho. 
Taking account of uncertainties in determination of the center of SNR Tycho, we regard a circular region with the radius of 30"  as the candidate region. 
A $V$-band image of  this region taken with Suprime-Cam is shown in Figure \ref{fig:2}. 
We designate the stars in the candidate region with capital alphabets same as Ruiz04 except Tycho(Y), 
which has no data in Ruiz04 (see Figure \ref{fig:2}). 
The stars out of the region are designated with lower case alphabets (see Figure \ref{fig:3}). 
We select 8 stars brighter than 22 mag ($V$) in this region for spectroscopic observations.  

\subsection{Spectroscopic Observations}
\subsubsection{Observations}
We carried out optical spectroscopy of stars in the region of SNR Tycho using multi-object-spectroscopy (MOS) mode 
of the Faint Object Camera and Spectrograph (FOCAS) (Kashikawa et al. 2002) attached to Subaru Telescope. 
One mask of FOCAS/MOS covers a 6' $\phi$ aperture diameter field of view. 
The MOS slit lengths were 8."0 and slit widths were 1."0.  
We used the 2nd order of the grism of 150 lines/mm and the L550 order-sorting filter. 
This setting gave wavelength coverage  from 3600 \AA\, to 5200 \AA\, and
the spectral resolution was R$\sim$400 at 3800 \AA. 
The pixel scale of the CCDs is 0."1 /pixel. 
We binned 3 pixels in the spatial direction. 
In order to get a higher throughput in short wavelength, observations were made without 
the atmospheric dispersion corrector (ADC). 

Observations were carried out on Sep. 14 and Sep. 15, 2006. 
We observed 8 stars in the candidate regions within 30" $\phi$ radius 
and other 25 stars located outside of the candidate region. 
The reason why we observed the outside stars is to confirm 
whether sufficient Fe I remains in SNR Tycho. 
The two brightest stars, (A) and (B) in the candidate region, are not regarded as candidates 
because these stars are so bright that they are very likely to be in the foreground of SNR Tycho. 
Seeing size during observations was $\sim$0."6 - 0."8. 
The exposure time of each frame was 0.5 hours 
and the total effective exposure time was 12.5 hours.
We dithered $\pm$0."2 for each exposure to avoid bad pixels. 
We made two MOS masks to observe all of the candidate stars. 
The bright 5 targets including Tycho(G) were observed using Mask1. 
The exposure time of Mask1 was 2.5 hours (0.5 hours $\times$ 5). 
Mask1 was used at low elevation because targets of Mask1 were relatively bright. 
The other 4 targets including Tycho(V) were observed using Mask2. 
The exposure time of Mask2 was 10.0 hours (0.5 hours $\times$ 20). 
The other 24 targets were observed using both masks. 

\vspace{1em}

\subsubsection{Data and reduction}
We analyze obtained spectra using IRAF software 
\footnote{IRAF is distributed by the National Optical Astronomy Observatory, 
which is operated by the Association of Universities for Research in Astronomy, Inc., 
under cooperative agreement with the National Science Foundation}.  
Subtracting the bias level and flatfielding are carried out. 
Then, each spectrum is extracted from multislit images. 
Each spectrum of 0.5 hour exposure time is combined with imcombine of IRAF task. 
At the same time, this task rejects cosmic rays and bad pixels above 3 $\sigma$. 
We carry out skysubtraction to use a 2nd order of spline3 function. 
We calibrate wavelengths using Th-Ar lamp. The one-dimensional spectra are made by apall of IRAF task and 
are flux-calibrated using spectra of a standard star, BD+28d4211, obtained with the long-slit spectroscopy 
mode of FOCAS with a 2."0 slit width.  
Then, we correct these spectra for the Galactic extinction and slit loss. 
This correction is described in the next section in detail.   
We analyze 4 spectra Tycho(C), (E), (F) and (G) in the candidate region 
and spectra of 13 stars outside the candidate region. 
The other spectra do not have good signal-to-noise ratios. 

Spectra of 21 stars before correction of the Galactic extinction and slit loss 
are available from the web page (http://www.ioa.s.u-tokyo.ac.jp/$^\sim$iharayt/SNRTycho/). 
Out of 21 stars, 8 stars are in the candidate region and 13 stars are outside the region. 
The above 21 spectra include 4 spectra of Tycho(D), (I), (V) and (Y) 
which do not have good signal-to-noise ratios ($<$3.0 at 3800 \AA) and we did not use them in this work. 
We analyzed the other 17 spectra with good signal-to-noise ratios ($>$3.0 at 3800\AA). 
They are shown in Figure 6 after the flux correction. 

\section{Method to detect Fe I absorption lines}

\subsection{Calibration of observed spectra}
We need to take into account two causes of the flux loss at blue wavelengths. 
One of them is the reddening due to the Galactic extinction. 
It is difficult to accurately estimate the Galactic extinction because of unknown distances to observed stars. 
The other is slit loss, which is expected to be a function of wavelength 
because we did not use ADC. 
We set the central wavelength of slit acquisition at 3900-4100 \AA.
The flux out of this wavelength range was gradually lost from the slit due to atmospheric dispersions. 
Atmospheric dispersion becomes larger at short wavelengths compared with that at long wavelengths. 
We have to handle these two complex flux losses at the same time. 
Therefore, we make flux corrections by multiplying a following linear function 
so that a template stellar spectrum traces the observed spectrum. 
\begin{equation}
f(\lambda)=1-a\times (c-\lambda)
\end{equation}
where $a$ is a variable and we correct the flux at shorter wavelengths than $c$. 
The values of $a$ are listed up in Table. 1. 
The value of $c$ is fixed to 3840 \AA, because we avoid the wavelength range of 3840-3880 \AA\,
where we expect Fe I absorption. 
   
Further details of spectral fitting and measuring possible absorption lines
are also described in \S 3.2 and \S 3.3.  

\subsection{Spectral types of observed spectra}
We compare the observed spectra with the template spectra by Jacoby, Hunter, \& Christian (1984) (hereafter Ja84). 
They incorporated spectra for many stars having spectral classes O-M and luminosity classes V, I\hspace{-.1em}I\hspace{-.1em}I, and I into their library. 
Their wavelength coverage is 3510-7400 \AA\, and the resolution is $\sim$4 \AA.  
Ja84 is the most suitable catalog compared with other spectral libraries, for example, 
Gunn \& Stryker 1983, Silva \& Cornell 1992, Soubiran, Katz \& Cayrel 1998, or Pickles 1998, 
in terms of the number of spectra, spectral resolution, and wavelength coverages. 

We determine spectral types by finding a template spectrum 
that most closely matches the observed spectrum of each star in the wavelength range of 3900-4400 \AA. 
We have chosen this wavelength range 
because Fe I absorption lines due to SNR are expected at wavelengths shorter than 3900 \AA, and 
also because the slit loss becomes large at around 4400 \AA\, and increases at longer wavelengths. 
The resolution of our observing mode was not good enough to distinguish 
whether the star was a dwarf or a giant. 
In other words, spectra of a dwarf and a giant are similar in low resolution. 
Thus we used only dwarf spectra for spectral fitting. 
This is sufficient for our aim to find possible absorption features of the SNR in observed spectra.
  
We first compare the observed spectra with some templates to see global line shapes at 3900-4400 \AA. 
Then we have chosen several template spectra by referring to the shapes of characteristic absorption lines 
such as the Ca H\&K lines at 3968 \AA\, and 3933 \AA, the H$\delta$ line at 4101 \AA, 
the CH line at 4300 \AA, and the H$\gamma$ line at 4340 \AA.
Finally, we calculate $\chi^2$ residuals of several candidate template spectra 
to determine the spectral types and the flux levels. 
To minimize possible errors in 3700-3900 \AA\, for detection of the Fe I absorptions,  
we determine the spectral type by choosing the template spectrum with the minimum $\chi^2$ residuals 
in the wavelength ranges of 3740-3840 \AA\, and 3900-4100 \AA. 
In this $\chi^2$ fitting, the template spectra are binned to match the resolution to 
that of the observed spectra. 
The spectral types thus determined are listed in Table 1. 
In this fitting, we do not consider radial velocities of stars because they are negligible compared with our spectral resolution.

\subsection{Measurement of absorption lines due to Fe I in the ejecta} 
\begin{figure}
  \begin{center}
    \FigureFile(80mm,80mm){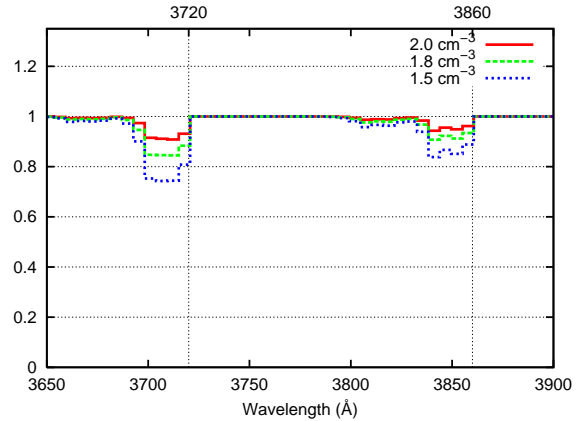}
  \end{center}
  \caption{The profiles of absorption lines in the spectrum of the companion star due to Fe I 
  in the SN ejecta predicted for different values of the density of the ambient medium 
  (Red solid line : $n_a=2.0$ cm$^{-3}$, Green dashed line: 1.8 cm$^{-3}$, Blue dotted line: 1.5 cm$^{-3}$). 
  The companion star is assumed to be located within 10" from the center of SNR Tycho. 
  }\label{fig:4}
\end{figure}

According to OS06, the spectrum of the companion star of Tycho's SN 
should show broad absorption lines of Fe I in the blue sides of 3720 \AA\, and  3860 \AA.
In addition, the same Fe I in the ejecta should imprint broad absorption lines 
on both of the blue and red sides of 3720 \AA\, and 3860 \AA\, 
in the spectra of background stars of SNR Tycho. 
Figure \ref{fig:4} shows the predicted profiles of these absorption lines of the companion star 
based on the W7 model (Nomoto, Thielemann, \& Yokoi 1984) for several densities $n_a$ of the ambient medium. 
The profile is sensitive to $n_a$ of SNR Tycho.  

In this calculation, the hydrodynamical code is revised from OS06. 
We adopt the original distribution of elements in W7 
instead of the simplified distribution used in OS06. 
As a result, elements such as Ca, Ar, S are added. 
Thus, we add new emission line data for the additional elements 
and revise line data of the rest elements. 
Moreover, the data set of collisional ionization cross section is updated.  
As we relax the assumption in OS06 that shock heated atoms are  
automatically ionized to some extent, more ionizing photons are  
emitted from atoms in lower ionization stages. As a result, for  
$n_a$=1.5-2.0, which is higher than that calculated in OS06, the  
expected absorption line profiles become shallower and weaker than  
those in OS06.

We measure equivalent widths of possible absorption lines. 
This equivalent width is different from the usually used equivalent width in the sense 
that the template spectrum is used for the reference instead of the continuum spectrum of the same star.
Therefore, we introduce the term "Excess Absorption Measure" (EAM).  
The EAM here is defined as
\begin{equation}
{\rm EAM}=\int\left(1-\frac{F_{\rm obs}(\lambda)}{F_{\rm template}(\lambda)}\right)d\lambda,
\end{equation}
where the observed flux is denoted by $F_{\rm obs}$ and the template flux $F_{\rm template}$.
We bin the spectra in order to reduce the photon noise and the readout noise. 
The binning size is 4 pixels (5.6 \AA\, per bin ). 

We consider two error sources. 
One is an error from the template fitting, and 
the other is a statistical error from photons and readout noise. 
The fitting error is estimated by root mean square values (R.M.S.) (=$E_{\rm fit}$) 
of the residuals in the wavelength ranges where we expected no absorption lines, i.e., 
in 3650-3680 \AA, 3740-3840 \AA\, and 3880-3920 \AA. 
The photon noise is estimated by the number of electrons originated 
from the stellar photons and the background photons at each wavelength, 
and signal-to-noise ratio (S/N) is given by the following formula,

\begin{eqnarray}
S/N &=& \frac{S}{\sqrt{S+3N+3\sigma_e^2}}, \\
E_{\rm  sta} &=& \frac{1}{S/N},
\end{eqnarray}
where $S$ is the number of photons of an object per 3 pixels, 
$N$ is the number of photons of the sky background per pixel 
averaged over adjacent 18 pixels, of which 9 pixels are on the right side of the signal and 9 pixels are on the left. 
$\sigma_e$ is the readout noise, which is 4 $e^-$ r.m.s/pixel for FOCAS. 
The gain is 2.1 $e^-$/ADU. 

We combine these two error sources and the total error $E_{\rm all}$ is estimated by the following formula, 
\begin{equation}
E_{\rm all} = \sqrt{E_{\rm fit}^2+E_{\rm sta}^2},
\end{equation}
assuming that the residuals of the fitting follow the Gaussian distribution. 
We plot the ratios of $E_{\rm all}$ to $F_{\rm obs}$ per bin in  Figure \ref{fig:c}-\ref{fig:am}. 

We calculate EAMs in five wavelength ranges, which are the blue and red sides of 3720 \AA\, and 3860 \AA\,, and around 3800 \AA. 
The EAMs are calculated by integrating 4 bins (22 \AA) which correspond to the absorption widths 
predicted for the blue wing of the companion star by OS06. 
The errors are also accumulated in the same bins. We summarize the results with errors in Table. 1. 

\section{Results}
\subsection{Spectral types}

\begin{table*}
\begin{center}
\rotatebox[origin=c]{90}{
\begin{minipage}{\textheight}
 \caption{Summary of observations }\label{tab:first}
    \begin{tabular}{c|c|c|c|c|c|c|c|c|c|c|c|c}
     Star name & Type & $V$ mag & Distance\footnotemark[a] & Mask\footnotemark[b] & S/N\footnotemark[c]& $\chi^2$\footnotemark[d] & a \footnotemark[d] & EAM(3720B)& EAM(3720R) & EAM(3800) & EAM(3860B) & EAM(3860R)\\
    \hline\hline
     Tycho(C) & M1V & 18.9 & 4.5" & 1\&2 & 7.5 & 6.33 & 0.0004 & 0.52$\pm$0.96 & -1.08$\pm$0.92 & 0.37$\pm$0.83 & 0.37$\pm$0.78 & 0.75$\pm$0.77 \\
     Tycho(D) &  -\footnotemark[$\ast$] & 20.6 & 5.6"& 1\&2 & 2.8 & - & - & - & - & - & - & - \\
     Tycho(E) & F8V & 19.7 & 10.8" & 1\&2 & 13.3 & 7.45 & 0.0006 & 3.00$\pm$0.55 & 0.56$\pm$0.52 & 0.54$\pm$0.44 & -0.10$\pm$0.42 & -0.30$\pm$0.41 \\
     Tycho(F) & F4V & 17.7 & 26.2" & 1    & 27.1 & 26.0 & 0.0023 &0.78$\pm$0.42 & 0.09$\pm$0.41 & 0.47$\pm$0.40 & 0.13$\pm$0.39 & -0.27$\pm$0.39 \\
     Tycho(G) & F8V & 18.7 & 26.2" & 1    & 14.9 & 11.9 & 0.0020 &0.59$\pm$0.53 & -0.25$\pm$0.51 & -0.36$\pm$0.45 & -0.06$\pm$0.43 & -0.58$\pm$0.42 \\
     Tycho(I) &  -  & 21.8 & 22.7" & 1\&2 & 2.3 & - & - &- & - & - & - & - \\
     Tycho(V) &  -  & 21.4 & 24.8" & 2    & 2.4 & - & - &- & - & - & - & - \\
     Tycho(Y) &  -  & 22.1 & 24.2" & 1\&2 & 2.0 &- & - &- & - & - & - & - \\
    \hline
     Tycho(aa) & F8V & 20.3 & 1.20' & 2   & 6.7 & 3.26 & 0.0028 & 3.06$\pm$1.07 & 0.26$\pm$0.96 & 0.51$\pm$0.71 & 0.14$\pm$0.61 & -0.61$\pm$0.58 \\
     Tycho(ab) & G6V & 20.8 & 1.26' & 1\&2& 3.1 & 1.39 & 0.0007 & 0.26$\pm$1.71 & 1.24$\pm$1.58 & 1.18$\pm$1.25 & 0.05$\pm$1.05 & 1.15$\pm$0.97 \\
     Tycho(ac) & F5V & 19.5 & 2.40' & 1   & 10.1 & 6.97 & 0.0023 & 0.61$\pm$0.86 & -1.11$\pm$0.80 & -0.04$\pm$0.67 & 0.82$\pm$0.62 & 0.53$\pm$0.60 \\
     Tycho(ad) & G3V & 20.9 & 1.27' & 2   & 3.9 & 1.45 & 0.0004 & 2.29$\pm$1.28 & 1.69$\pm$1.20 & 0.06$\pm$1.03 & -0.24$\pm$0.93 & 0.24$\pm$0.89 \\
     Tycho(ae) & G6V & 20.1 & 2.01' & 1\&2& 6.7 & 2.06 & 0.0012 &-0.24$\pm$0.86 & 0.67$\pm$0.82 & 0.21$\pm$0.72 & 0.94$\pm$0.67 & 0.89$\pm$0.65 \\
     Tycho(af) & G2V & 20.3 & 2.20' & 1\&2& 7.3 & 3.36 & 0.0004 &-0.91$\pm$1.11 & -0.53$\pm$1.01 & 0.80$\pm$0.83 & -0.28$\pm$0.77 & 0.18$\pm$0.75 \\
     Tycho(ag) & F6V & 19.6 & 1.00' & 1\&2& 16.3 & 6.47 & 0.0015 & 0.20$\pm$0.41 & -0.74$\pm$0.39 & 0.17$\pm$0.34 & 0.56$\pm$0.31 & 0.02$\pm$0.30 \\
     Tycho(ah) & F8V & 19.7 & 1.94' & 1\&2& 13.2 & 8.92 & 0.0022 & 2.91$\pm$0.42 & 0.72$\pm$0.39 & 0.10$\pm$0.33 & -0.01$\pm$0.29 & -0.53$\pm$0.28 \\
     Tycho(ai) & G3V & 21.0 & 1.42' & 1\&2& 3.8 & 2.06 & 0.0016 & 3.86$\pm$1.94 & 2.22$\pm$1.81 & -0.45$\pm$1.56 & 2.06$\pm$1.49 & 1.86$\pm$1.47 \\
     Tycho(aj) & F8V & 20.3 & 2.38' & 1\&2& 8.8 & 4.53 & 0.0023 & -0.58$\pm$0.84 & 0.01$\pm$0.79 & -0.06$\pm$0.70 & 0.31$\pm$0.66 & -0.57$\pm$0.65 \\
     Tycho(ak) & G3V & 20.9 & 1.85' & 1\&2& 3.9 & 1.16 & 0.0004 & 2.67$\pm$1.60 & 1.85$\pm$1.50 & -1.21$\pm$1.24 & 1.14$\pm$1.11 & 1.76$\pm$1.06 \\
     Tycho(al) & G6V & 20.1 & 2.27' & 1\&2& 6.8 & 2.26 & 0.0008 & 0.52$\pm$0.97 & 0.62$\pm$0.91 & -0.21$\pm$0.77 & 0.13$\pm$0.70 & -0.13$\pm$0.68 \\
     Tycho(am) & F7V & 19.9 & 3.74' & 1\&2& 12.3 & 6.56 & 0.0007 & 0.54$\pm$0.64 & -0.03$\pm$0.60 & -0.37$\pm$0.54 & -0.10$\pm$0.50 & 0.62$\pm$0.49 \\
    \hline
    \end{tabular}
  \end{minipage}
  }
  \rotatebox[origin=c]{90}{
  \begin{minipage}{\textheight}
  \footnotemark[a] The distance is the projected distance from the center of SNR Tycho
   which is ($\alpha$, $\delta$) = (\timeform{00h25m19.40s}, \timeform{64D08'13''.98}) (J2000.0).  \\
  \footnotemark[b] We made two MOS masks according to stellar positions (see \S 2.2.1). \hspace{2em}
  \footnotemark[c] S/Ns are measured at 3800 \AA\, per 1 pixel (1.4 \AA). \\
  \footnotemark[d] They are fitting parameters. Their details are written in \S3.1 and \S3.2. \\ 
  \footnotemark[$\ast$] The stellar types of Tycho(D), (I), (V) and (Y) are unknown for not good signal-to-noise ratios. 
  \end{minipage}
  }
  \end{center}
\end{table*}

\begin{figure*}
  \begin{minipage}{0.5\hsize}
  \begin{center}
    \FigureFile(80mm,50mm){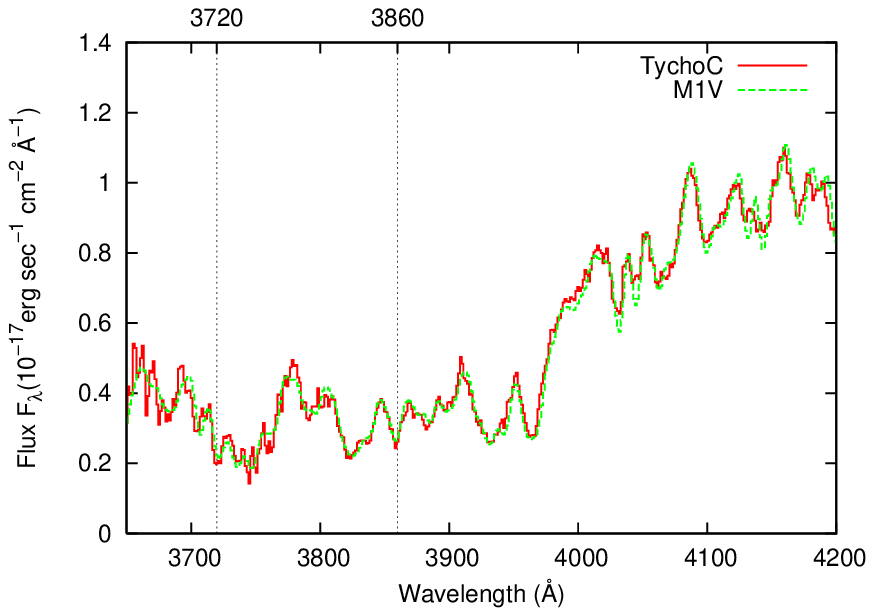}
  \end{center}
  \center{{\footnotesize {\bf Fig. \ref{fig:c}a.} Tycho(C)}}
  \end{minipage}
  \begin{minipage}{0.5\hsize}
  \begin{center}
    \FigureFile(80mm,50mm){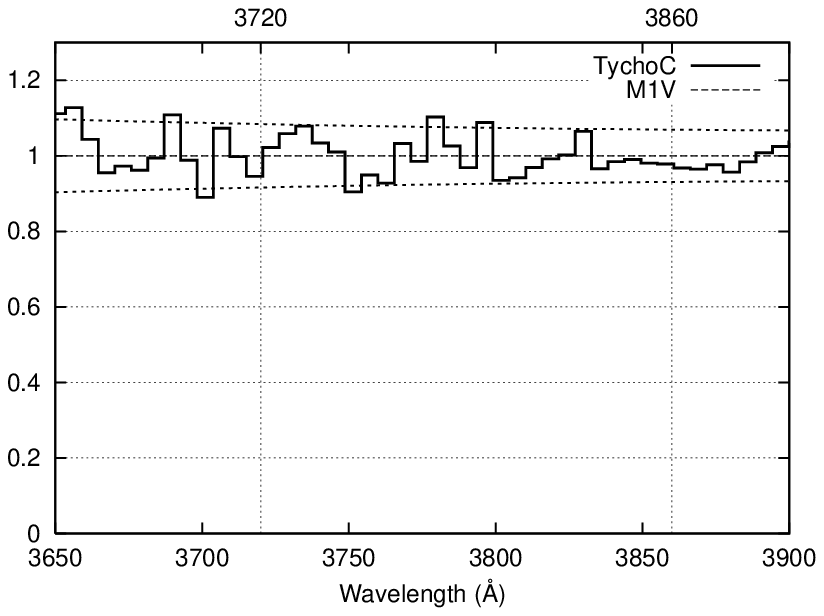}
  \end{center}
  \center{{\footnotesize {\bf Fig. \ref{fig:c}b.} Tycho(C), ratio}}
  \end{minipage}
  \caption{Comparisons of the observed spectra with template stellar spectra (Jacoby, Hunter, \& Christian 1984). 
  The left panels are our observed spectra (Red solid lines) and the fitted template spectra (Green dashed lines). 
  The right panels show the ratio of observed spectra to the fitted template spectra. 
  The dashed lines show 1$\sigma$ error per bin estimated by (5). 
  According to Ozaki \& Shigeyama (2006), absorption lines of Fe I are expected 
  at $\sim$3720 \AA\, and $\sim$3860 \AA\,
  if stars are within or behind SNR Tycho, 
  and those wavelengths are shown by two dotted vertical lines.}
  \label{fig:c}
\end{figure*}

\begin{figure*}
  \begin{minipage}{0.5\hsize}
  \begin{center}
    \FigureFile(80mm,50mm){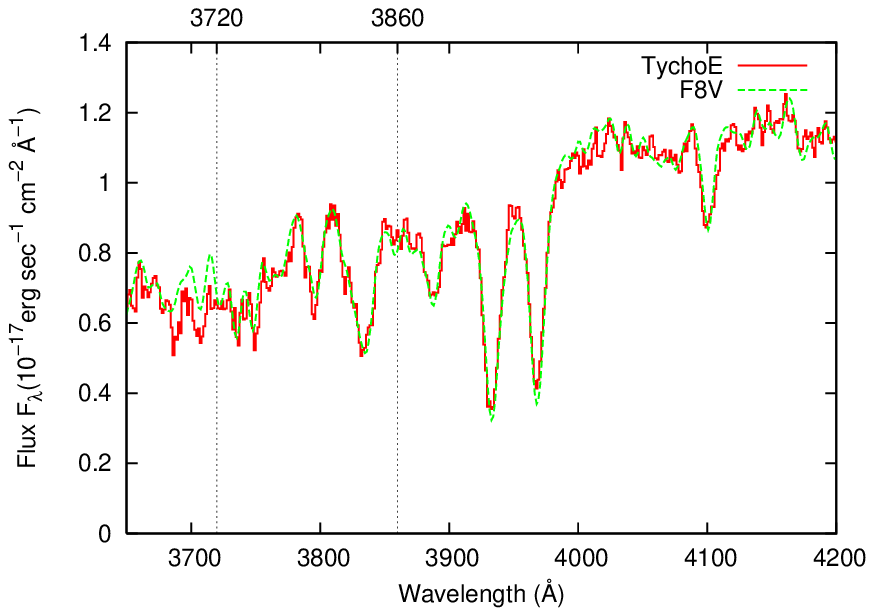}
  \end{center}
  \center{{\footnotesize {\bf Fig. \ref{fig:e}a.} Tycho(E)}}
  \end{minipage}
  \begin{minipage}{0.5\hsize}
  \begin{center}
    \FigureFile(80mm,50mm){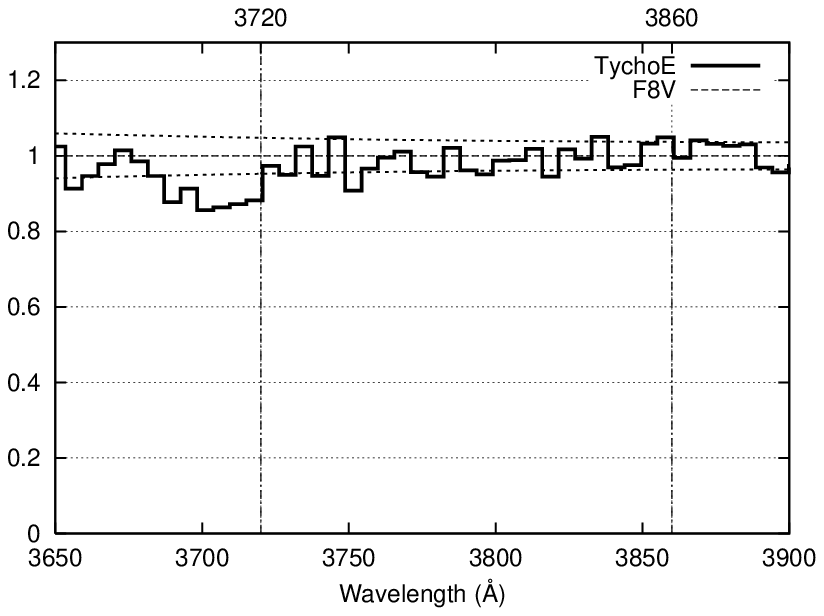}
  \end{center}
  \center{{\footnotesize {\bf Fig. \ref{fig:e}b.} Tycho(E), ratio}}
  \end{minipage}
  \caption{Same as Figure \ref{fig:c}, but for Tycho(E).}
  \label{fig:e}
\end{figure*}

\begin{figure*}
  \begin{minipage}{0.5\hsize}
  \begin{center}
    \FigureFile(80mm,50mm){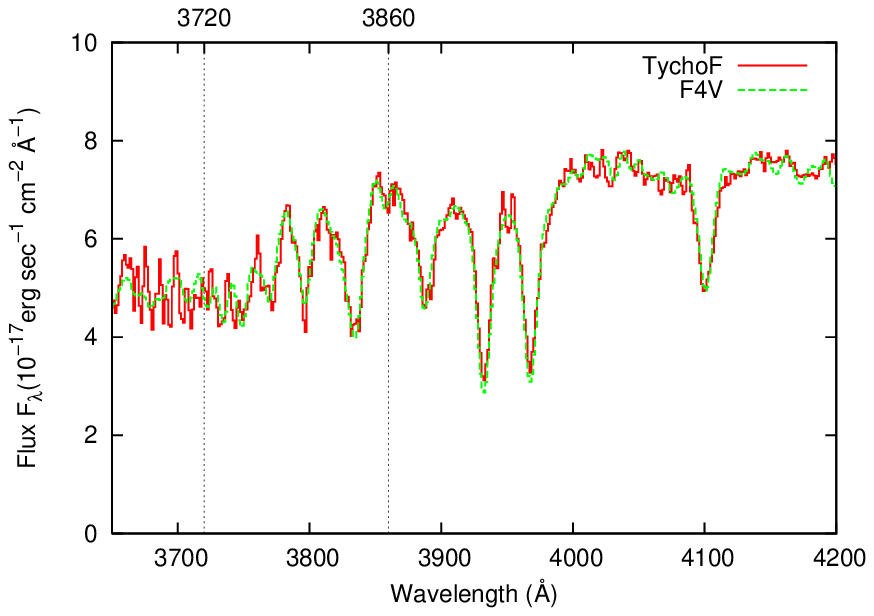}
  \end{center}
  \center{{\footnotesize {\bf Fig. \ref{fig:f}a.} Tycho(F)}}
  \end{minipage}
  \begin{minipage}{0.5\hsize}
  \begin{center}
    \FigureFile(80mm,50mm){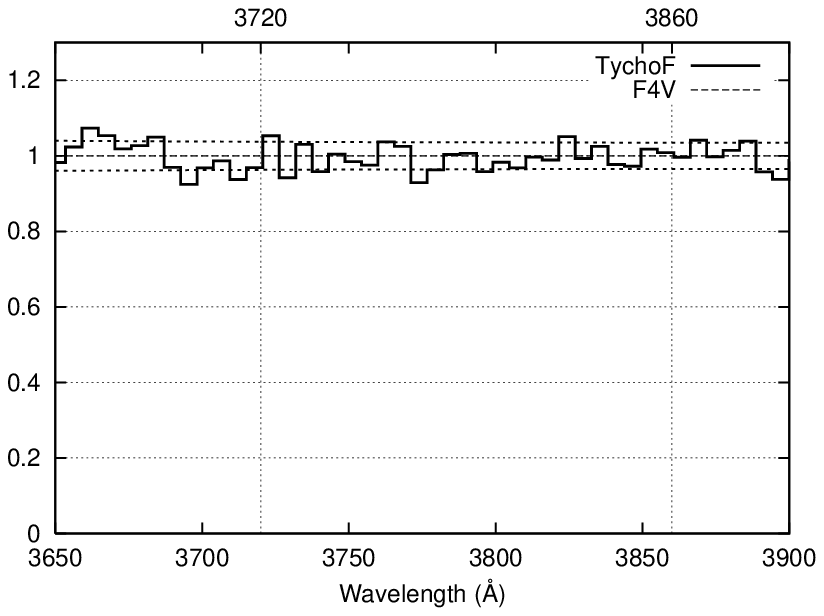}
  \end{center}
  \center{{\footnotesize {\bf Fig. \ref{fig:f}b.} Tycho(F), ratio}}
  \end{minipage}
  \caption{Same as Figure \ref{fig:c}, but for Tycho(F).}
  \label{fig:f}
\end{figure*}

\begin{figure*}
  \begin{minipage}{0.5\hsize}
  \begin{center}
    \FigureFile(80mm,50mm){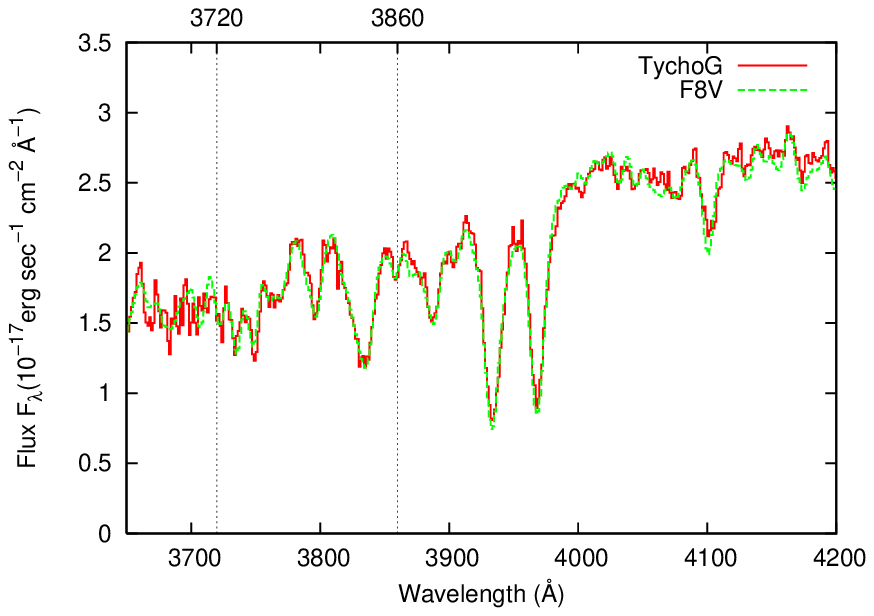}
  \end{center}
  \center{{\footnotesize {\bf Fig. \ref{fig:g}a.} Tycho(G)}}
  \end{minipage}
  \begin{minipage}{0.5\hsize}
  \begin{center}
    \FigureFile(80mm,50mm){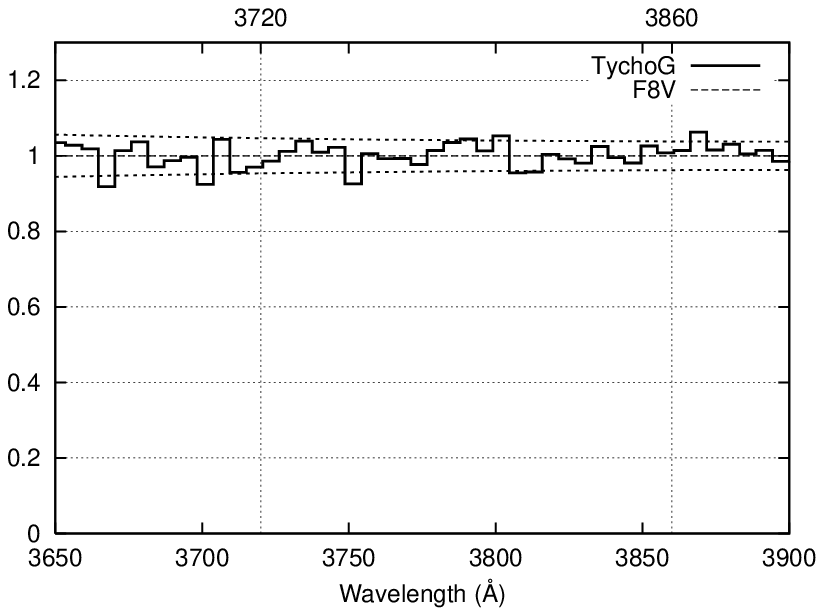}
  \end{center}
  \center{{\footnotesize {\bf Fig. \ref{fig:g}b.} Tycho(G), ratio}}
  \end{minipage}
  \caption{Same as Figure \ref{fig:c}, but for Tycho(G).}
  \label{fig:g}
\end{figure*}

\begin{figure*}
  \begin{minipage}{0.5\hsize}
  \begin{center}
    \FigureFile(80mm,50mm){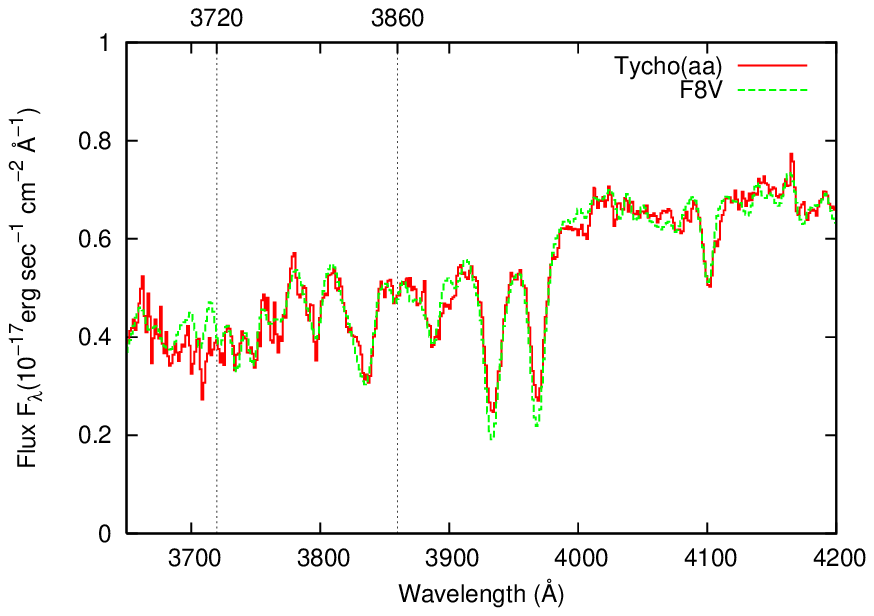}
  \end{center}
  \center{{\footnotesize {\bf Fig. \ref{fig:aa}a.} Tycho(aa)}}
  \end{minipage}
  \begin{minipage}{0.5\hsize}
  \begin{center}
    \FigureFile(80mm,50mm){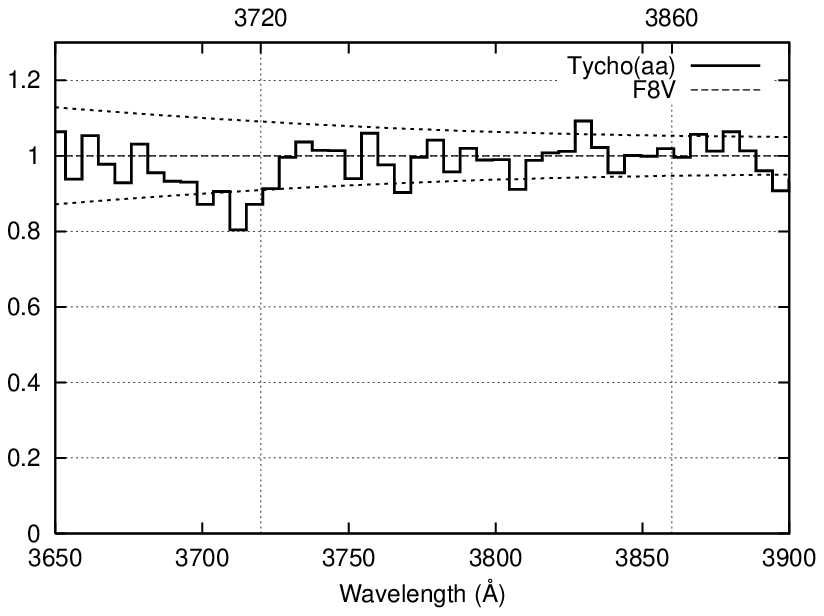}
  \end{center}
  \center{{\footnotesize {\bf Fig. \ref{fig:aa}b.} Tycho(aa), ratio}}
  \end{minipage}
  \caption{Same as Figure \ref{fig:c}, but for Tycho(aa).}
  \label{fig:aa}
\end{figure*}

\begin{figure*}
  \begin{minipage}{0.5\hsize}
  \begin{center}
    \FigureFile(80mm,50mm){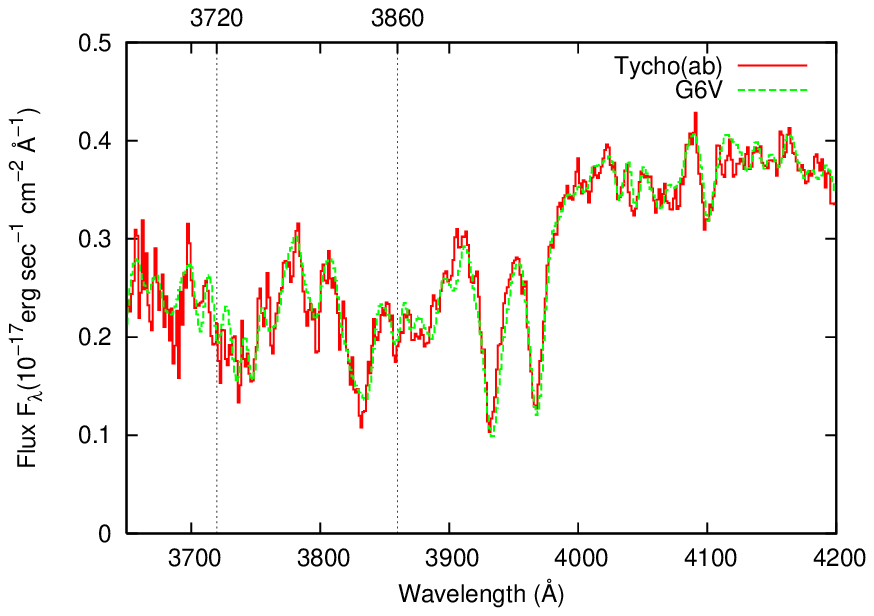}
  \end{center}
  \center{{\footnotesize {\bf Fig. \ref{fig:ab}a.} Tycho(ab)}}
  \end{minipage}
  \begin{minipage}{0.5\hsize}
  \begin{center}
    \FigureFile(80mm,50mm){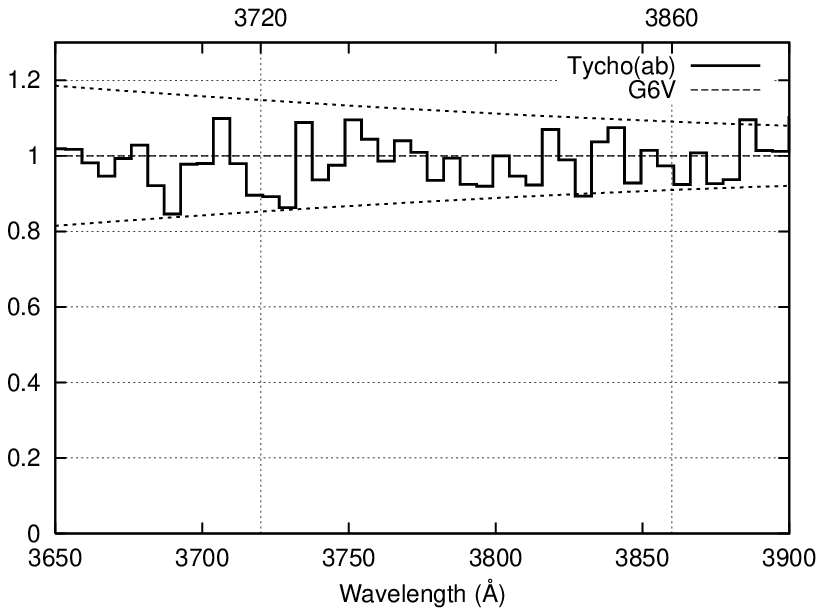}
  \end{center}
  \center{{\footnotesize {\bf Fig. \ref{fig:ab}b.} Tycho(ab), ratio}}
  \end{minipage}
  \caption{Same as Figure \ref{fig:c}, but for Tycho(ab).}
  \label{fig:ab}
\end{figure*}

\begin{figure*}
  \begin{minipage}{0.5\hsize}
  \begin{center}
    \FigureFile(80mm,50mm){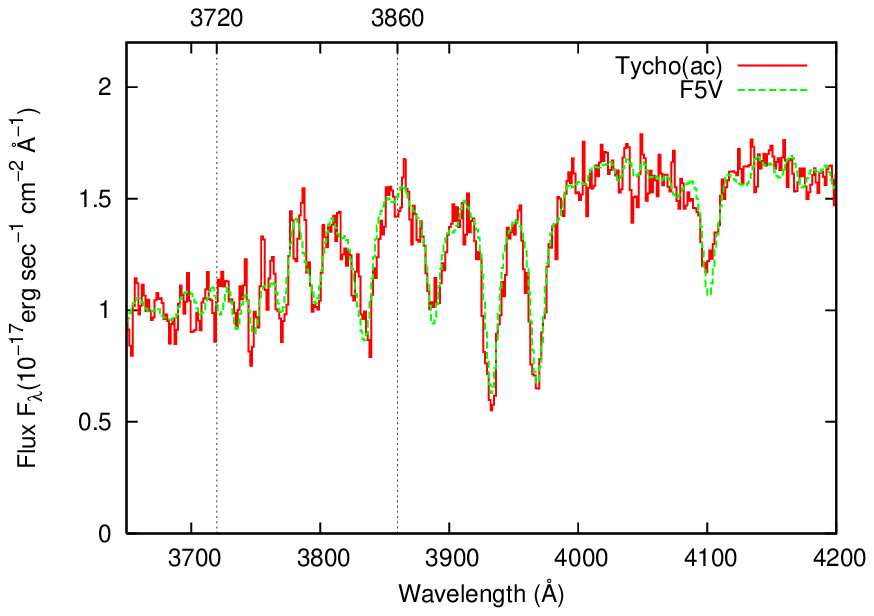}
  \end{center}
  \center{{\footnotesize {\bf Fig. \ref{fig:ac}a.} Tycho(ac)}}
  \end{minipage}
  \begin{minipage}{0.5\hsize}
  \begin{center}
    \FigureFile(80mm,50mm){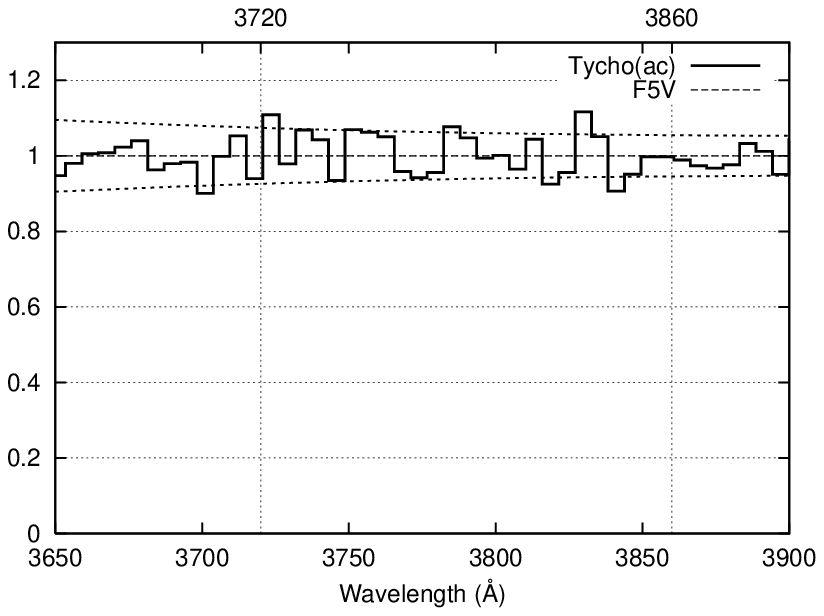}
  \end{center}
  \center{{\footnotesize {\bf Fig. \ref{fig:ac}b.} Tycho(ac), ratio}}
  \end{minipage}
  \caption{Same as Figure \ref{fig:c}, but for Tycho(ac).}
  \label{fig:ac}
\end{figure*}

\begin{figure*}
  \begin{minipage}{0.5\hsize}
  \begin{center}
    \FigureFile(80mm,50mm){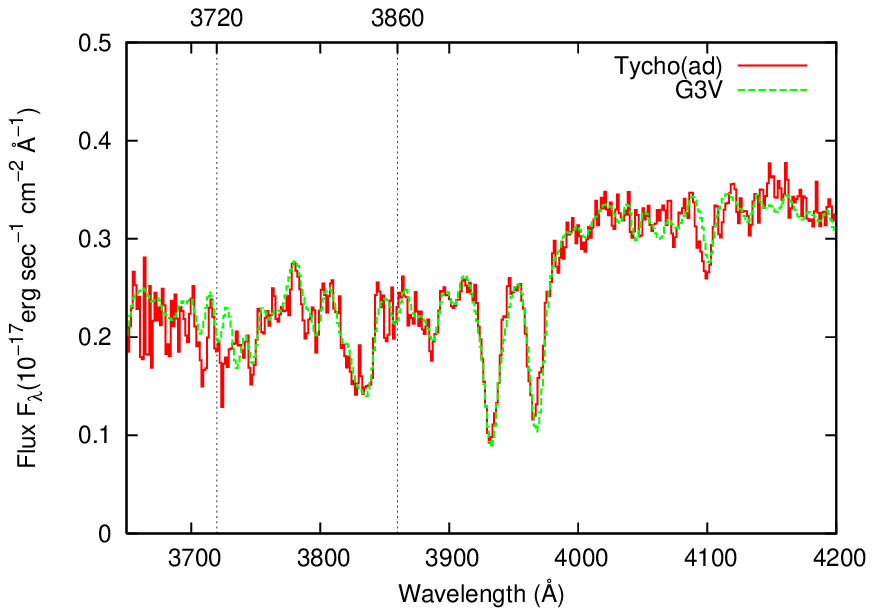}
  \end{center}
  \center{{\footnotesize {\bf Fig. \ref{fig:ad}a.} Tycho(ad)}}
  \end{minipage}
  \begin{minipage}{0.5\hsize}
  \begin{center}
    \FigureFile(80mm,50mm){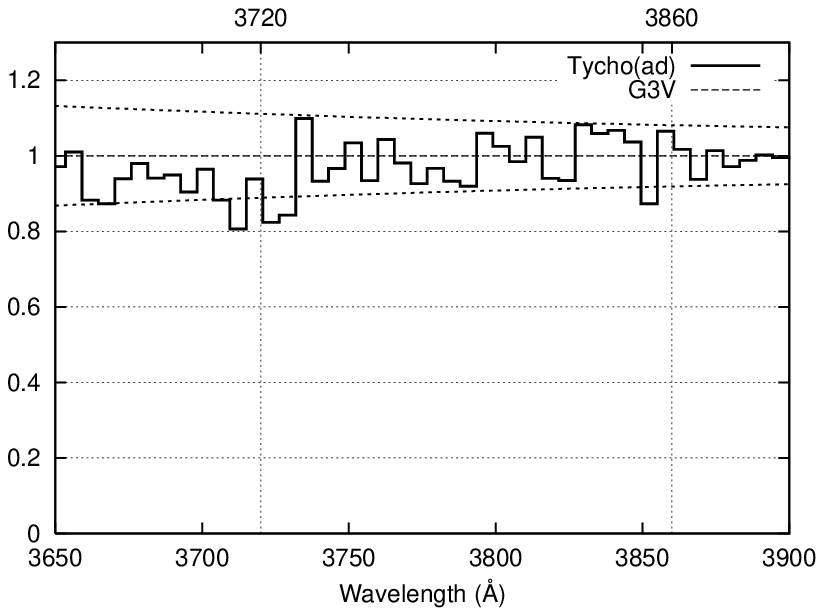}
  \end{center}
  \center{{\footnotesize {\bf Fig. \ref{fig:ad}b.} Tycho(ad), ratio}}
  \end{minipage}
  \caption{Same as Figure \ref{fig:c}, but for Tycho(ad).}
  \label{fig:ad}
\end{figure*}

\begin{figure*}
  \begin{minipage}{0.5\hsize}
  \begin{center}
    \FigureFile(80mm,50mm){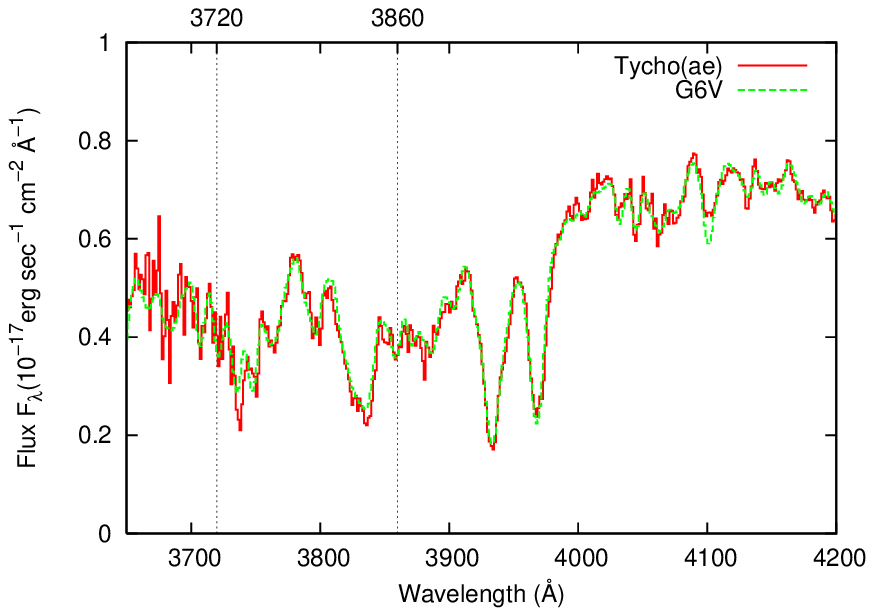}
  \end{center}
  \center{{\footnotesize {\bf Fig. \ref{fig:ae}a.} Tycho(ae)}}
  \end{minipage}
  \begin{minipage}{0.5\hsize}
  \begin{center}
    \FigureFile(80mm,50mm){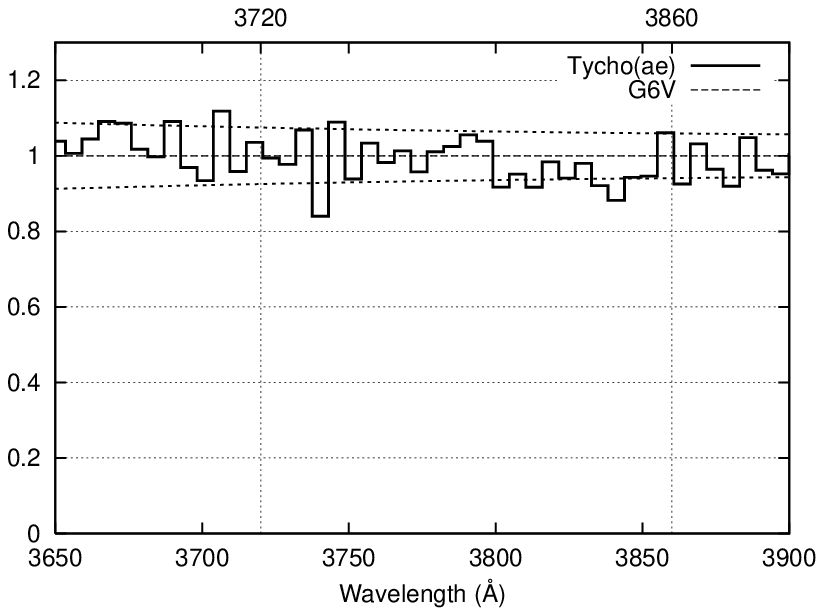}
  \end{center}
  \center{{\footnotesize {\bf Fig. \ref{fig:ae}b.} Tycho(ae), ratio}}
  \end{minipage}
  \caption{Same as Figure \ref{fig:c}, but for Tycho(ae).}
  \label{fig:ae}
\end{figure*}

\begin{figure*}
  \begin{minipage}{0.5\hsize}
  \begin{center}
    \FigureFile(80mm,50mm){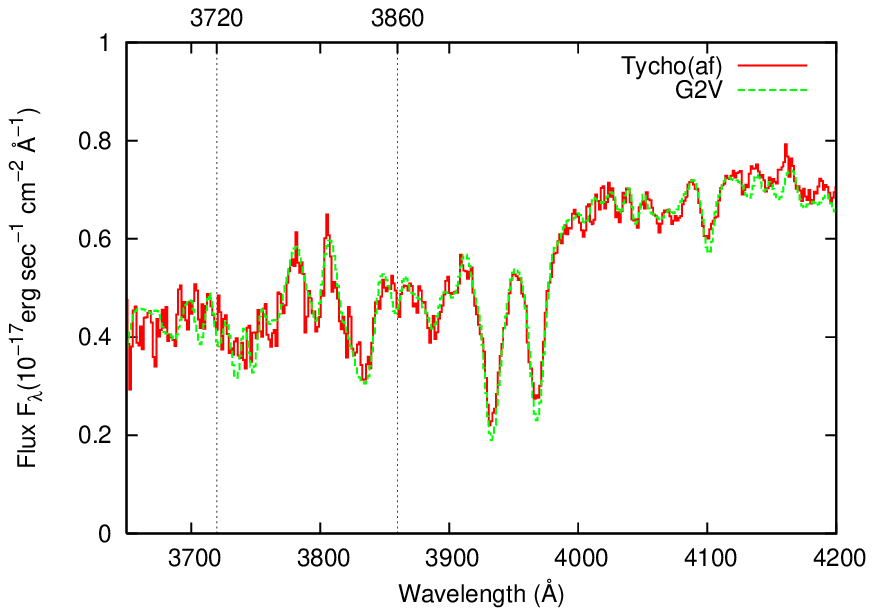}
  \end{center}
  \center{{\footnotesize {\bf Fig. \ref{fig:af}a.} Tycho(af)}}
  \end{minipage}
  \begin{minipage}{0.5\hsize}
  \begin{center}
    \FigureFile(80mm,50mm){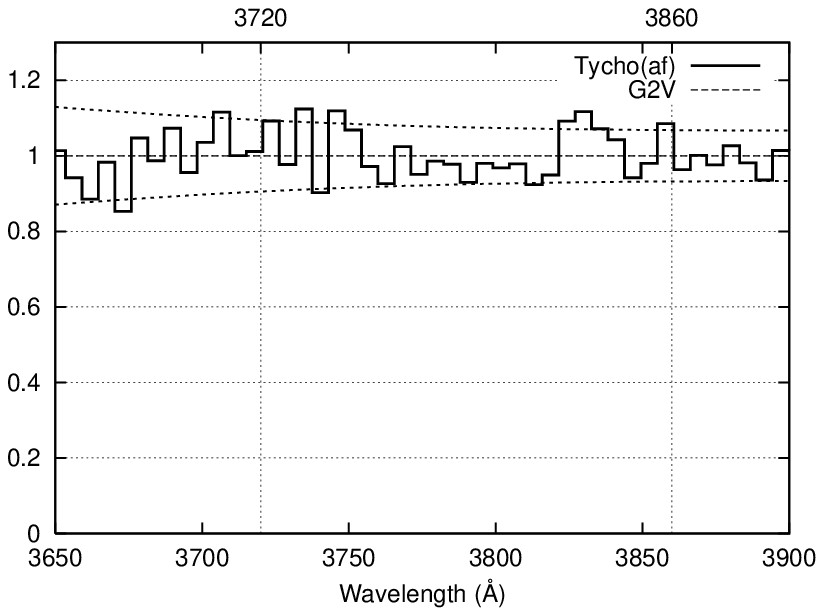}
  \end{center}
  \center{{\footnotesize {\bf Fig. \ref{fig:af}b.} Tycho(af), ratio}}
  \end{minipage}
  \caption{Same as Figure \ref{fig:c}, but for Tycho(af).}
  \label{fig:af}
\end{figure*}

\begin{figure*}
  \begin{minipage}{0.5\hsize}
  \begin{center}
    \FigureFile(80mm,50mm){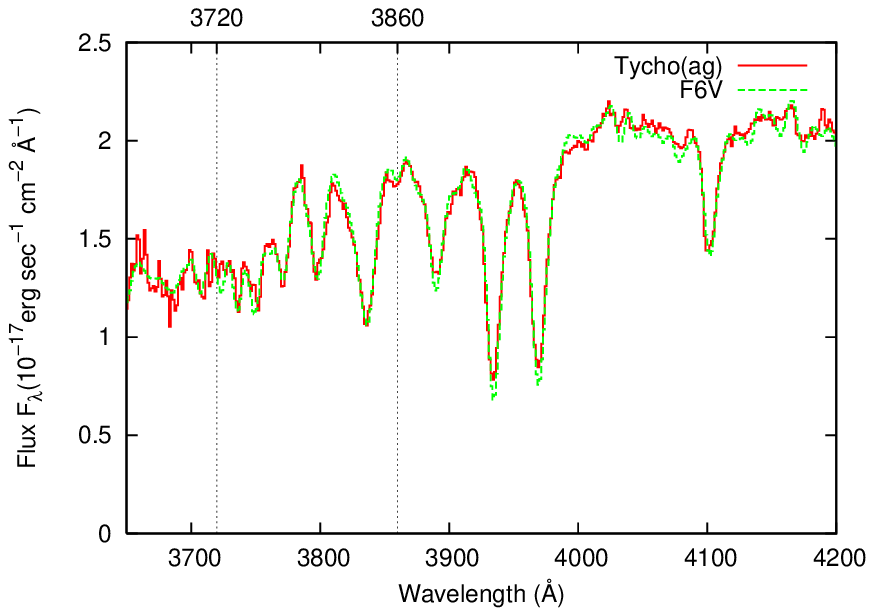}
  \end{center}
  \center{{\footnotesize {\bf Fig. \ref{fig:ag}a.} Tycho(ag)}}
  \end{minipage}
  \begin{minipage}{0.5\hsize}
  \begin{center}
    \FigureFile(80mm,50mm){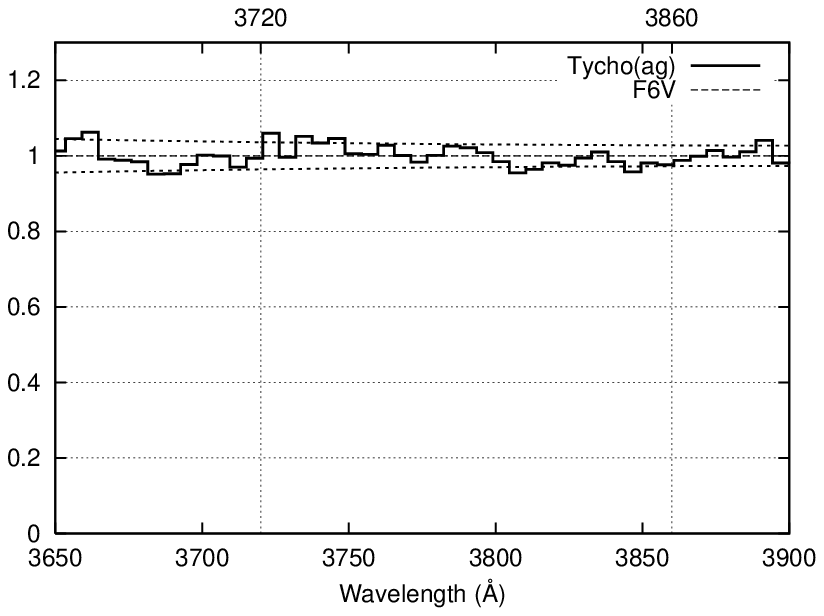}
  \end{center}
  \center{{\footnotesize {\bf Fig. \ref{fig:ag}b.} Tycho(ag), ratio}}
  \end{minipage}
  \caption{Same as Figure \ref{fig:c}, but for Tycho(ag).}
  \label{fig:ag}
\end{figure*}

\begin{figure*}
  \begin{minipage}{0.5\hsize}
  \begin{center}
    \FigureFile(80mm,50mm){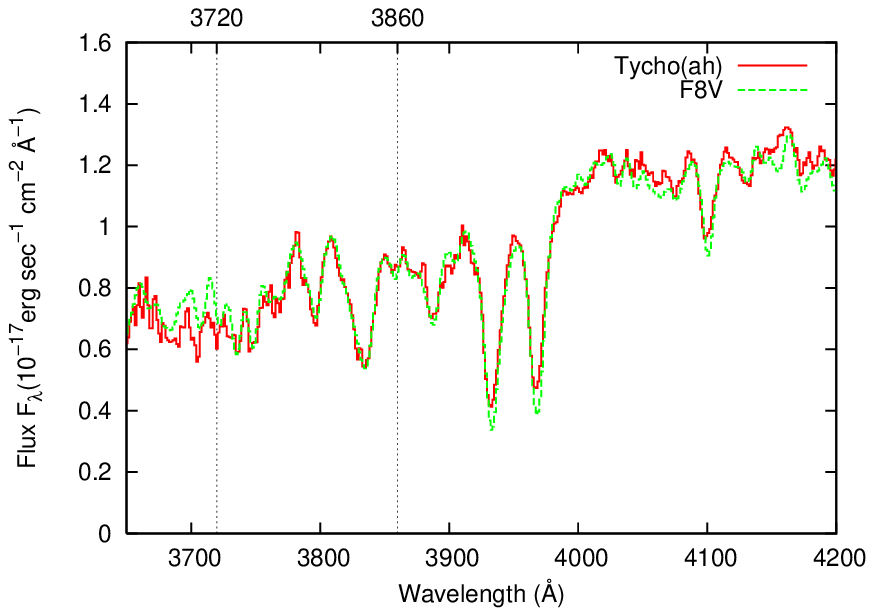}
  \end{center}
  \center{{\footnotesize {\bf Fig. \ref{fig:ah}a.} Tycho(ah)}}
  \end{minipage}
  \begin{minipage}{0.5\hsize}
  \begin{center}
    \FigureFile(80mm,50mm){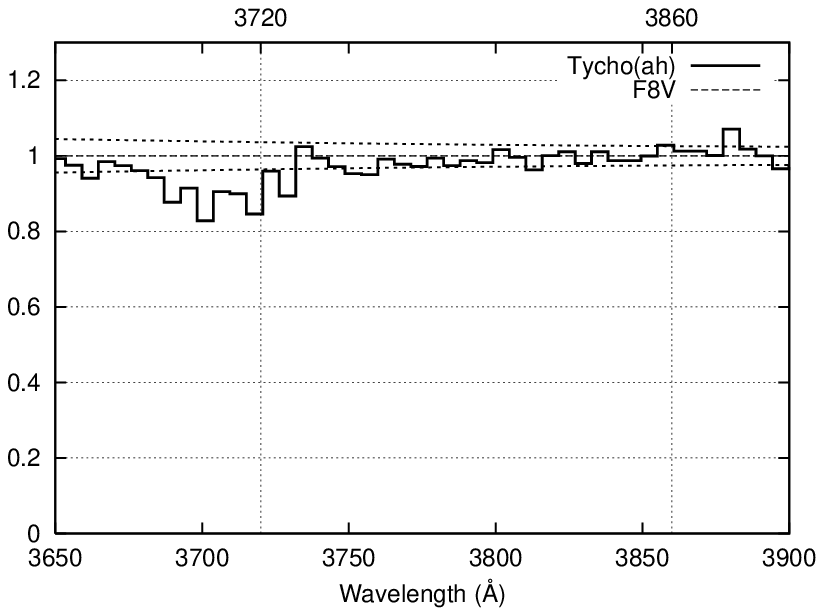}
  \end{center}
  \center{{\footnotesize {\bf Fig. \ref{fig:ah}b.} Tycho(ah), ratio}}
  \end{minipage}
  \caption{Same as Figure \ref{fig:c}, but for Tycho(ah).}
  \label{fig:ah}
\end{figure*}

\begin{figure*}
  \begin{minipage}{0.5\hsize}
  \begin{center}
    \FigureFile(80mm,50mm){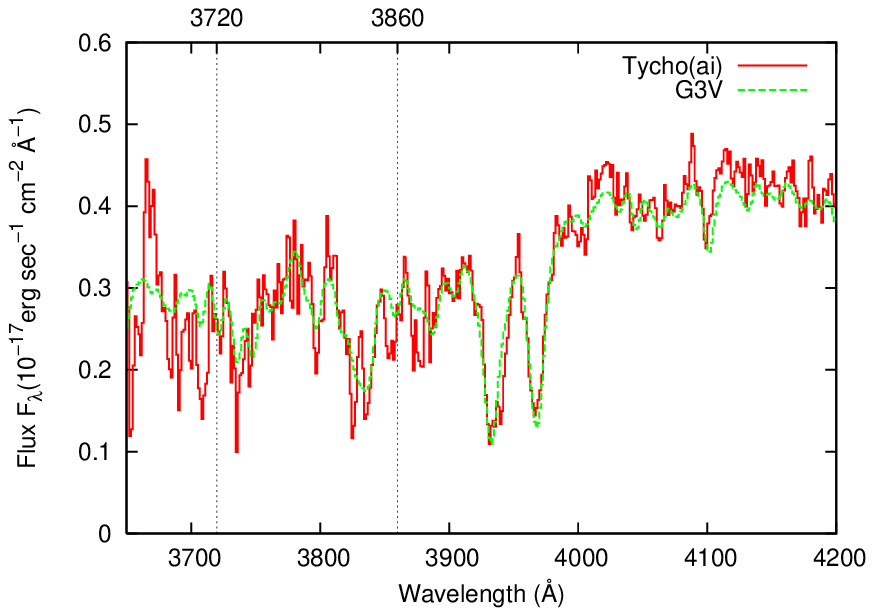}
  \end{center}
  \center{{\footnotesize {\bf Fig. \ref{fig:ai}a.} Tycho(ai)}}
  \end{minipage}
  \begin{minipage}{0.5\hsize}
  \begin{center}
    \FigureFile(80mm,50mm){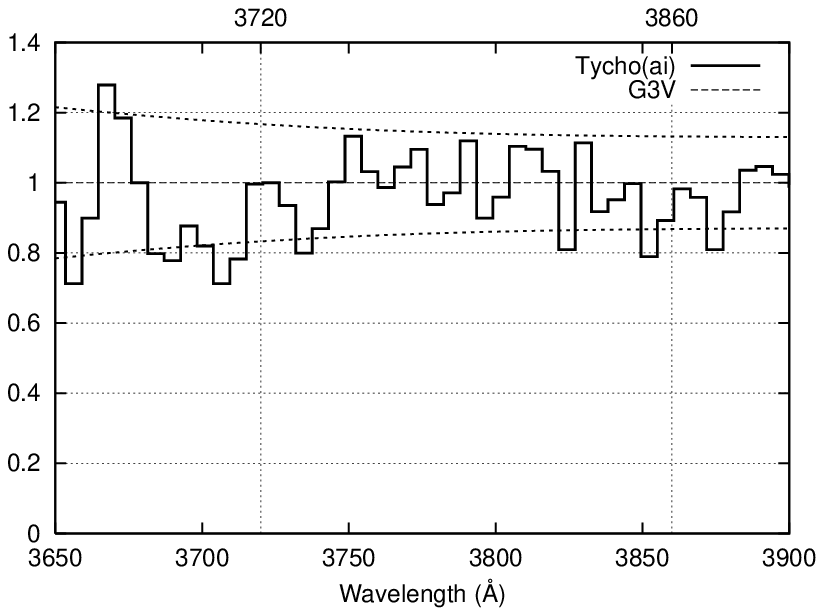}
  \end{center}
  \center{{\footnotesize {\bf Fig. \ref{fig:ai}b.} Tycho(ai), ratio}}
  \end{minipage}
  \caption{Same as Figure \ref{fig:c}, but for Tycho(ai).}
  \label{fig:ai}
\end{figure*}

\begin{figure*}
  \begin{minipage}{0.5\hsize}
  \begin{center}
    \FigureFile(80mm,50mm){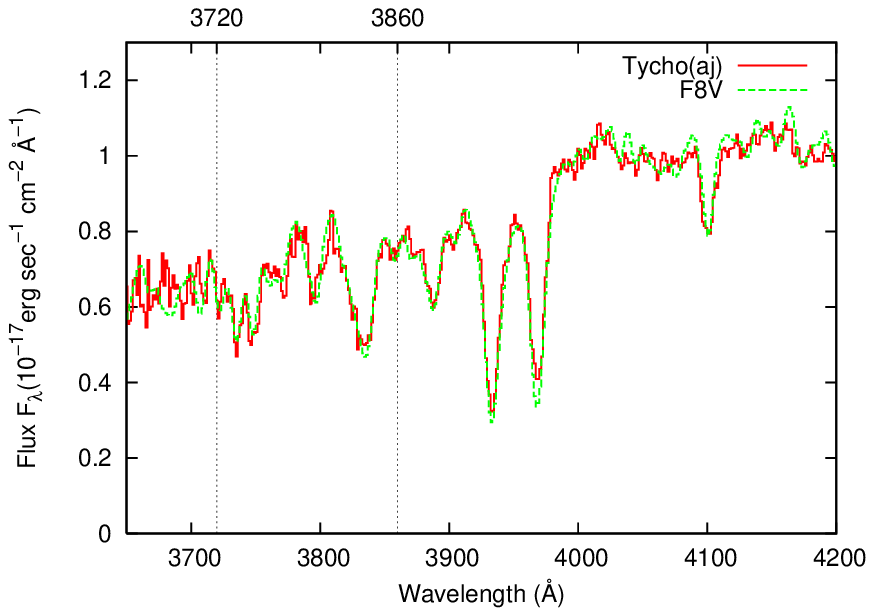}
  \end{center}
  \center{{\footnotesize {\bf Fig. \ref{fig:aj}a.} Tycho(aj)}}
  \end{minipage}
  \begin{minipage}{0.5\hsize}
  \begin{center}
    \FigureFile(80mm,50mm){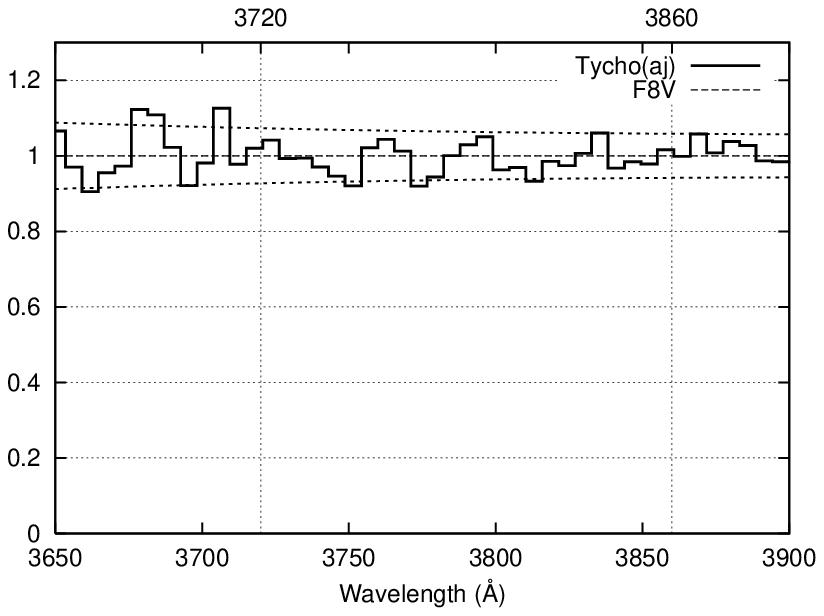}
  \end{center}
  \center{{\footnotesize {\bf Fig. \ref{fig:aj}b.} Tycho(aj), ratio}}
  \end{minipage}
  \caption{Same as Figure \ref{fig:c}, but for Tycho(aj).}
  \label{fig:aj}
\end{figure*}
  
\begin{figure*}
  \begin{minipage}{0.5\hsize}
  \begin{center}
    \FigureFile(80mm,50mm){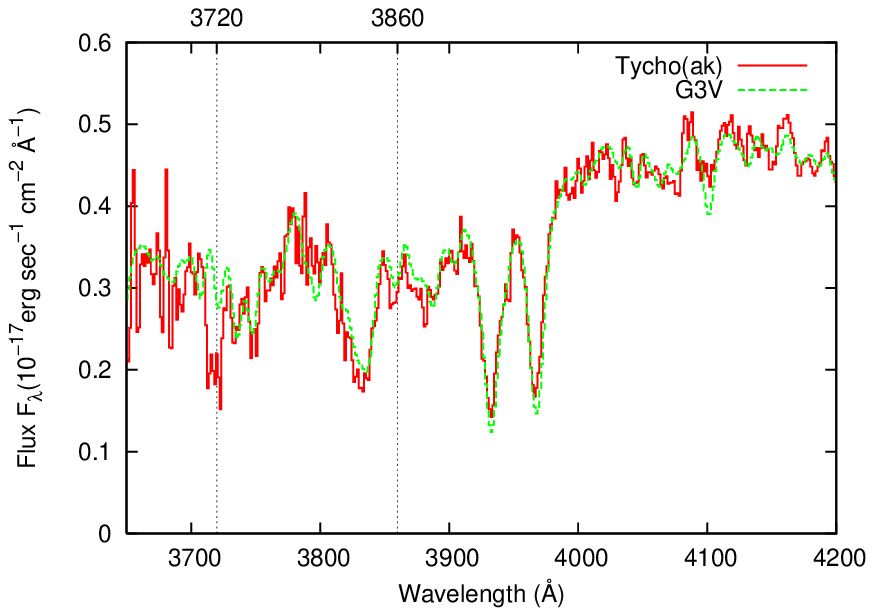}
  \end{center}
  \center{{\footnotesize {\bf Fig. \ref{fig:ak}a.} Tycho(ak)}}
  \end{minipage}
  \begin{minipage}{0.5\hsize}
  \begin{center}
    \FigureFile(80mm,50mm){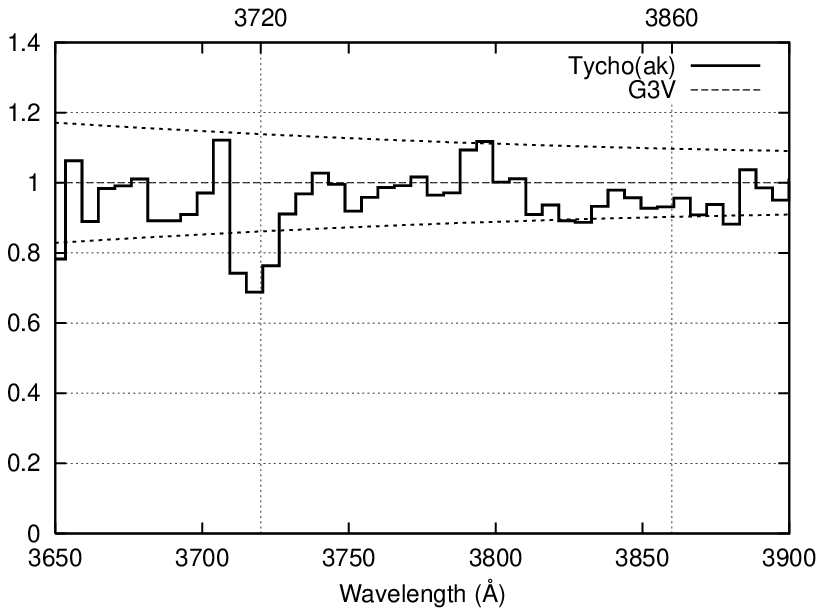}
  \end{center}
  \center{{\footnotesize {\bf Fig. \ref{fig:ak}b.} Tycho(ak), ratio}}
  \end{minipage}
  \caption{Same as Figure \ref{fig:c}, but for Tycho(ak).}
  \label{fig:ak}
\end{figure*}

\begin{figure*}[htb]
  \begin{minipage}{0.5\hsize}
  \begin{center}
    \FigureFile(80mm,50mm){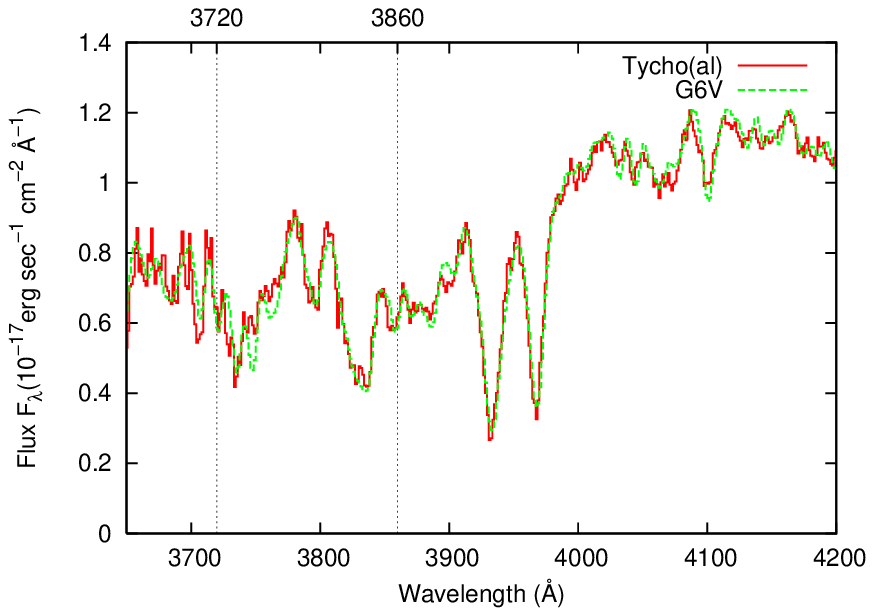}
  \end{center}
  \center{{\footnotesize {\bf Fig. \ref{fig:al}a.} Tycho(al)}}
  \end{minipage}
  \begin{minipage}{0.5\hsize}
  \begin{center}
    \FigureFile(80mm,50mm){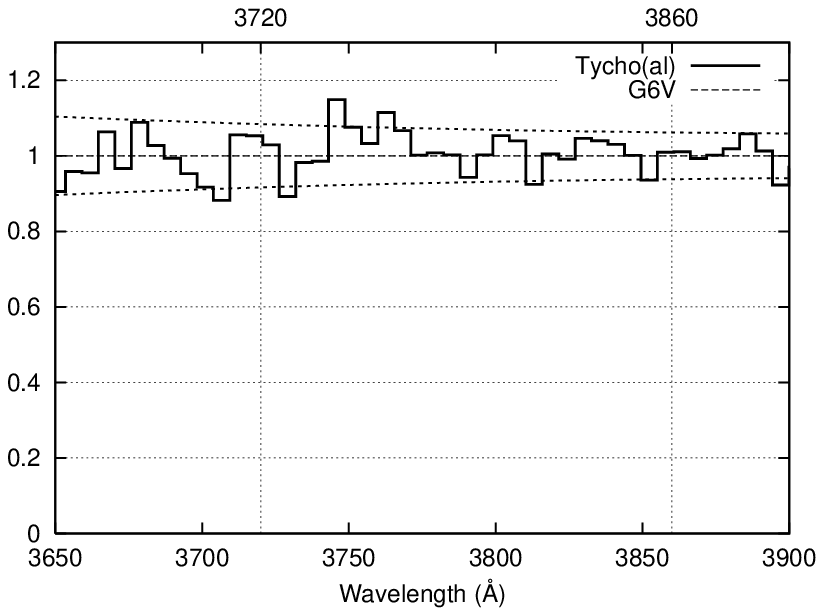}
  \end{center}
  \center{{\footnotesize {\bf Fig. \ref{fig:al}b.} Tycho(al), ratio}}
  \end{minipage}
  \caption{Same as Figure \ref{fig:c}, but for Tycho(al).}
  \label{fig:al}
\end{figure*}

\begin{figure*}
  \begin{minipage}{0.5\hsize}
  \begin{center}
    \FigureFile(80mm,50mm){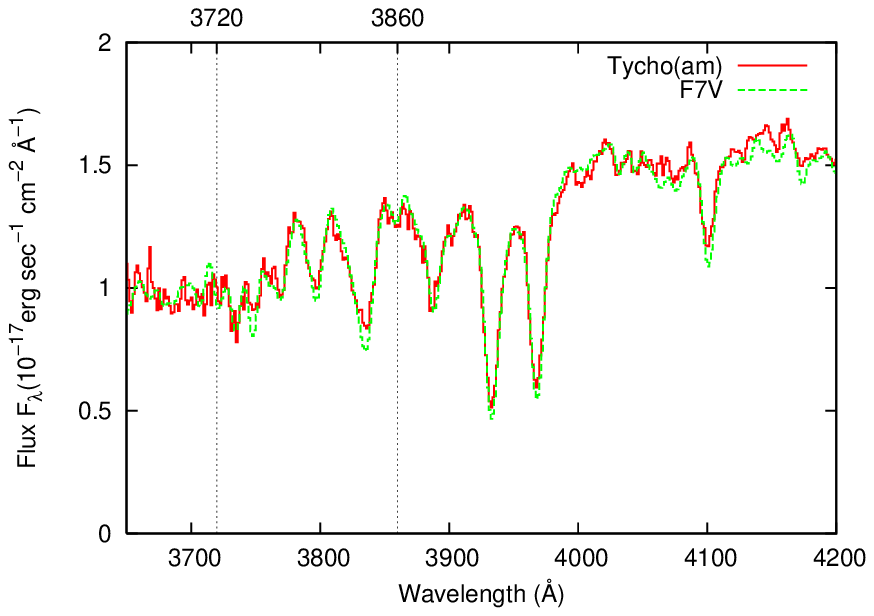}
  \end{center}
  \center{{\footnotesize {\bf Fig. \ref{fig:am}a.} Tycho(am)}}
  \end{minipage}
  \begin{minipage}{0.5\hsize}
  \begin{center}
    \FigureFile(80mm,50mm){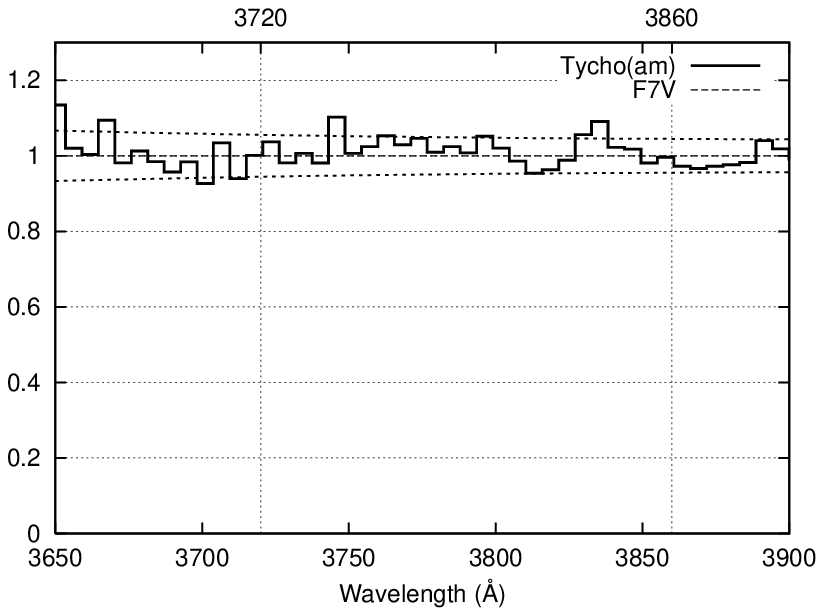}
  \end{center}
  \center{{\footnotesize {\bf Fig. \ref{fig:am}b.} Tycho(am), ratio}}
  \end{minipage}
  \caption{Same as Figure \ref{fig:c}, but for Tycho(am).}
  \label{fig:am}
\end{figure*}

The results are summarized in Figures \ref{fig:c}-\ref{fig:am} and Table 1.
In Table 1, we list names of stars, fitted spectral types, the number of masks used in the spectroscopy, 
signal-to-noise ratios, fitting parameters ($\chi^2$ and $a$), 
magnitudes in V-band, projected distances from the center of the SNR, 
and EAMs (of the blue and red wings of lines at 3720 \AA, at 3860 \AA, and 3790-3810 \AA). 
The EAM of 3790-3810 \AA\, (labeled as 3800 in Table 1) is used to see 
how the template spectra match the observed spectra. 
Although we take account of only main sequence stars in the spectral fitting, 
spectral types from this method are different from the results of Ruiz04. 
For example, we classify Tycho(G) as a F8V star while Ruiz04 classified it as a G2IV star. 
Furthermore, we classify Tycho(E) as a F8V star, while Ruiz04 classified it as a K2-3I\hspace{-.1em}I\hspace{-.1em}I star, 
and we classify Tycho(F) as a F4V star, while Ruiz04 classified it as a F9I\hspace{-.1em}I\hspace{-.1em}I star. 
Ruiz04 classified stellar types using the spectra at around 6500 \AA.  
On the other hand, we obtained spectra with lower resolutions in the range of 3900-4400 \AA\, (see \S3.2). 

\subsection{Absorption feature}
\subsubsection{Tycho(G)}
Since Tycho(G) was raised as the companion star of Tycho's SN by Ruiz04, we discuss this star first. 
As shown in Figures \ref{fig:g} and Table 1, 
the observed spectrum of Tycho(G) does not exhibit any significant broad absorption lines 
(EAM(3720B)$=0.59\pm0.53 \AA$) expected due to Fe I in the ejecta 
and is consistent with the template spectrum of a  F8V star. 
Therefore we do not obtain supporting evidence for Tycho(G) as the companion star of Tycho's SN. 
Nevertheless, this result does not refute the association of this star with Tycho SNR. 
There might remain only a fraction of Fe I in the ejecta insufficient 
to make broad absorption lines in the spectrum of the companion star. 

\subsubsection{Tycho(E)}
\begin{figure}
  \begin{center}
    \FigureFile(80mm,80mm){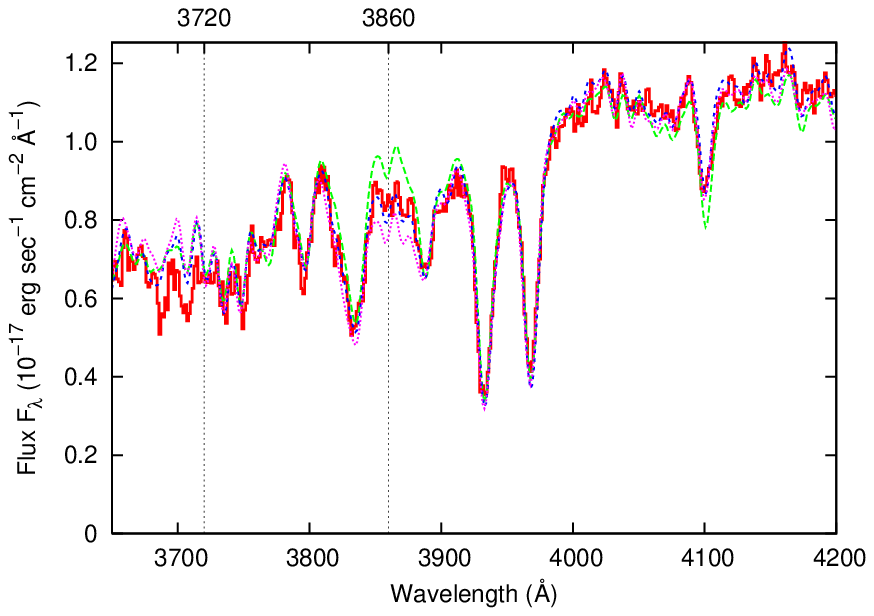}
  \end{center}
  \begin{center}
    \FigureFile(80mm,80mm){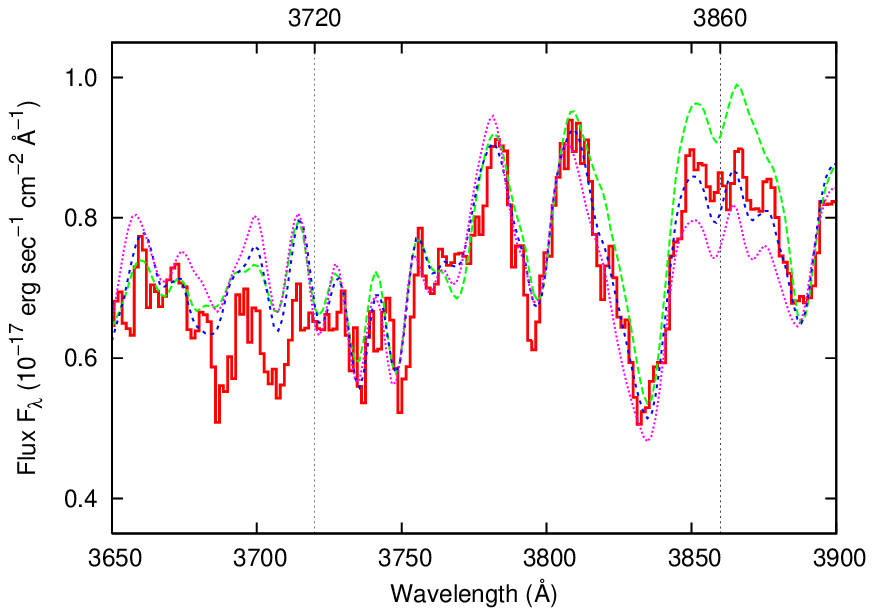}
  \end{center}
  \caption{Comparisons of the observed spectrum of Tycho(E) (red solid curve) 
  with the template spectra of F7V (green dashed curve), F8V (blue dashed curve), 
  and F9V (purple dashed curve) stars of Ja84. 
  The upper panel shows spectra in the whole observed range of wavelength. 
  The bottom panel is a close-up of the spectra in a limited wavelength range 3650-3900 \AA.
  The three template stars show similar behavior at around 3720 \AA, while 
  they are significantly different at around 3860 \AA. 
  Besides, a large difference is not seen at the other wavelengths. 
  }\label{fig:6}
\end{figure}

Significant broad absorption lines with positive EAMs are detected from stars Tycho (E), (aa), (ad), (ah), (ai), and (ak). Since only Tycho(E) is a candidate star for the companion of Tycho's SN in terms of its position, 
we take this star as an example and discuss the absorption profile of the spectrum. 
First, we discuss the spectrum around 3720 \AA.
There is a significant absorption with the EAM of 3.0 \AA\, (5.5 $\sigma$) in the blue side of 3720 \AA. 
The profile of the absorption resembles the theoretical prediction with $n_a$=1.8 cm$^{-3}$ in Figure \ref{fig:4}. 
The absorption line appears to extend to 3685 \AA. 
On the other hand, no significant absorption is detected around 3860 \AA. 
If the absorption at 3720 \AA\, is due to Fe I, 
the absorption with about a half of the EAM at 3720 \AA\, is expected at 3860 \AA. 
Since the accuracy of fitting around 3860 \AA\, is rather poor, 
we also fit the observed spectrum with spectra of F7V and F9V stars in addition to F8V (Figure \ref{fig:6}) 
to examine the influence of the adopted spectral type upon detections of broad absorption lines. 
As shown in Figure \ref{fig:6}, the shapes of the three spectra at around 3860 \AA\, differ from each other, 
while the shapes at around 3720 \AA\, are similar.  
If Tycho(E) had a spectrum of a F7V star as its original spectrum, 
the EAM in the blue side of 3860 \AA\, would become large ($\sim$2.5 \AA). 
However, the observed spectrum is fitted with F8V better than F7V. 
If we fitted with the middle type between F7V and F8V, we may detect some absorption at 3860 \AA. 
Because of a large variation of stellar spectra at around 3860 \AA, 
it is difficult to quantitatively discuss the relatively weak absorption line at 3860 \AA.

Ruiz04 classified this star as K2-3I\hspace{-.1em}I\hspace{-.1em}I 
and estimated the distance greater than 20 kpc. 
This apparently contradicts our classification based on the spectrum observed in 3900-4400 \AA. 
The spectrum of a catalogued star of K2I\hspace{-.1em}I\hspace{-.1em}I (or K3I\hspace{-.1em}I\hspace{-.1em}I) is quite different from that of F8V 
not only in this wavelength range but also at around 6500 \AA\, 
that Ruiz04 used to deduce the spectral type. 
As shown in the color-color diagram of the observed stars (Fig. \ref{fig:1a}),
the colors of Tycho(E) are $V - R_\mathrm{c} =0.9$ and  
$V - I_\mathrm{c} =1.9$, similar to that of Tycho(G).  
Both of the stars are classified as F8V. 
The colors of Tycho(C), which is classified as M1V, are $V - R_\mathrm{c} =1.2$ 
and $V - I_\mathrm{c} =2.4$.
If a K2-3I\hspace{-.1em}I\hspace{-.1em}I star is located at the distance of 20kpc,   
the colors of K2-3I\hspace{-.1em}I\hspace{-.1em}I should be $V - R_\mathrm{c}>1.2$ and $V - I_\mathrm{c}>2.5$ 
because this star is affected by larger the Galactic extinction 
than other stars which are located at the distance of 3kpc or less.   
As shown in the color-color diagram, 
stars are located on a sequence that has some scatter  
in the region with $V-R_\mathrm{c}\ge 3$. Tycho(E) and  
Tycho(G) are located left on the sequence while the K2-3I\hspace{-.1em}I\hspace{-.1em}I star  
should be located right on the sequence because a M1V star Tycho(C)  
is located just on the sequence.
Therefore, the classification of Tycho(E) by Ruiz04 contradicts the colors of Tycho(E). 
It is not at all impossible that Tycho(E) might have a quite peculiar spectrum 
if taking these results at face value. 
Further spectroscopic observations covering whole of the above wavelengths 
are needed to decide the spectral type of this star.

\subsubsection{Tycho(ah)}\label{tychoh}
\begin{figure}
  \begin{center}
    \FigureFile(80mm,80mm){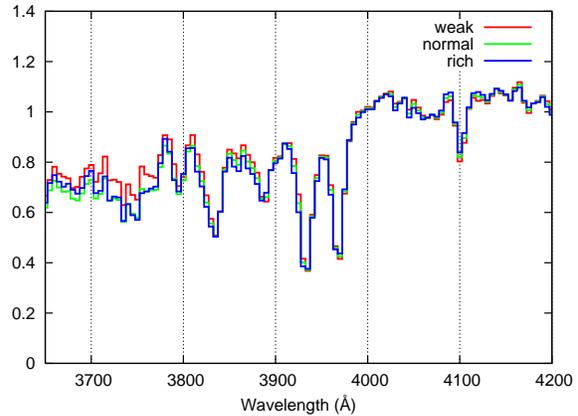}
  \end{center}
  \caption{Comparison of the template spectra of  F8V stars with different metallicities from Pickels 1998. 
  The red line shows the spectrum of a metal-weak F8V star, 
  the green line shows a F8V star with the solar metallicity, and the blue line shows a metal-rich F8V star. 
  The green line almost follows the spectrum of the F8V star of Ja84. 
  The difference between the metal-rich and the normal star does not affect 
  the amount of the absorption at $\sim$3720 \AA. 
  If we use the metal-weak star, the amount of the absorption is overestimated. 
  }\label{fig:f8v}
\end{figure}

\begin{figure}
  \begin{center}
    \FigureFile(80mm,80mm){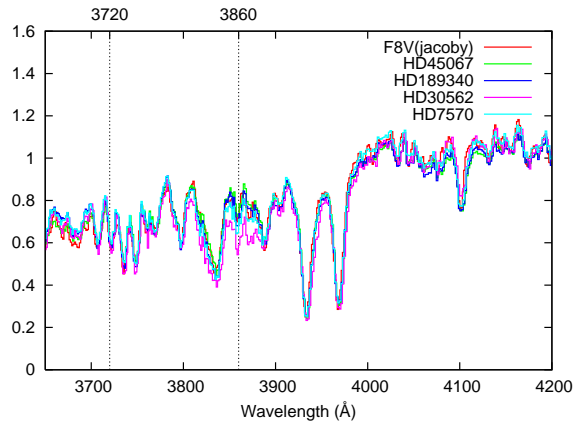}
  \end{center}
  \caption{Comparison of the template spectra of F8V stars in Bagnulo et al (2003). 
  These spectra were obtained by members of the ESO Paranal Science Operations Team 
  in order to make a library of high-resolution spectra of stars across the 
  Hertzsprung-Russell diagram. 
  The spectra by Bagnulo et al (2003) are converted into lower resolution spectra suitable for Ja84.
  We plot the four F8V stars in the catalog and one F8V star in Ja84 (red line). 
  None of their spectra can exhibit the absorption feature like Tycho(E) and (ah). 
  }\label{fig:eso}
\end{figure}

The absorption lines are seen at 3720 \AA\, from the spectra of other stars, 
but most of their profiles are different from that of Tycho(E). 
Tycho(ah) has a broad absorption line with the profile very similar to that of Tycho(E). 
In fact, the overall spectra of these two stars are quite similar.
The projected distance of Tycho(ah) from the geometrical center of SNR Tycho is  $\sim$2'. 
Thus if Tycho(E) is close to the SN site, then Tycho(ah) should have no physical interaction with the SN and vice versa.
If Tycho(ah) is one of the background stars of the SNR, 
the spectrum should have a broad absorption line in the red side of $\sim$3720 \AA\, 
in addition to the blue side. Our observations indicate that this would not be the case. 
Hence, the observed spectra of Tycho(E) and (ah) may indicate 
that they belong to a peculiar stellar type of F8V that has some deficit at around $\sim$3720 \AA\, 
compared with the F8V of Ja84.
This deficit might be caused by metal richness of the star as the spectra of F8V stars 
with different metallicities in the catalogue of Pickels (1998) shown in Figure \ref{fig:f8v}. 
We also compare with the F8V stars in the catalogue of Bagnulo et al (2003) shown in Figure \ref{fig:eso}. 
A lot of metal absorption lines from 3720 \AA\, to 3900 \AA\, changing the shape of the spectrum 
in this region make it difficult to deduce the accurate EAMs of the blue wing of 3720 \AA\, and 3860 \AA\, absorption.  
To do this, we need to observe various nearby stars with the spectral type of F8V 
toward the direction of SNR Tycho and out of the region of SNR Tycho. 


\section{Discussion}
\begin{figure*}
  \begin{minipage}{0.5\hsize}
  \begin{center}
    \FigureFile(80mm,50mm){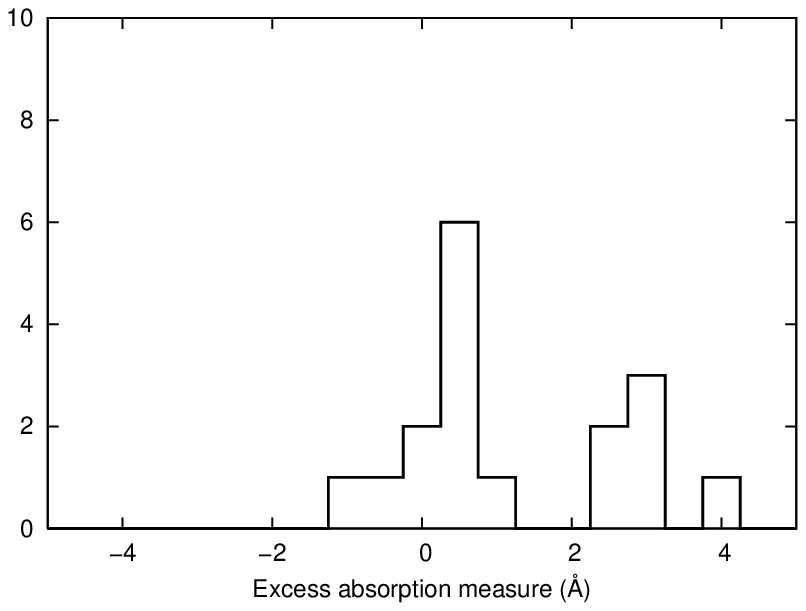}
  \end{center}
  \center{{\footnotesize {\bf Fig. \ref{fig:7}a.} Distribution of EAM at 3700-3720 \AA}}
  \end{minipage}
  \begin{minipage}{0.5\hsize}
  \begin{center}
    \FigureFile(80mm,80mm){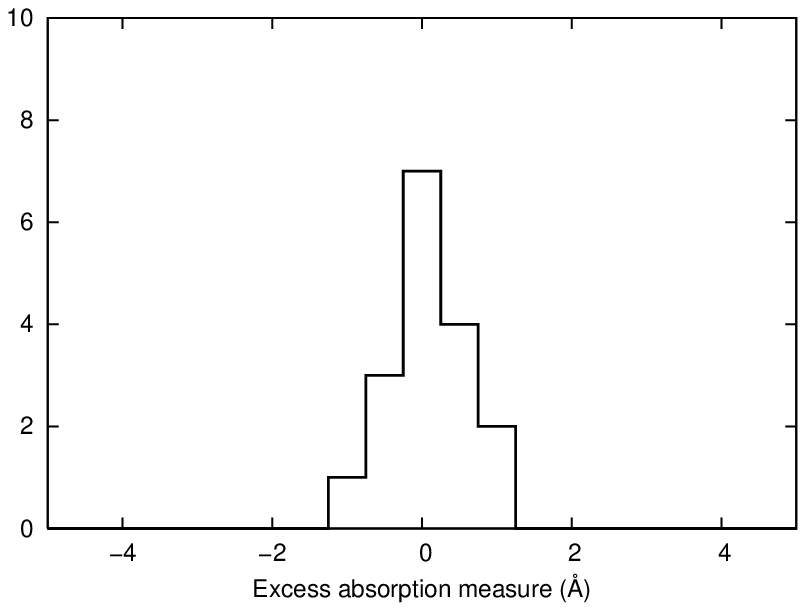}
  \end{center}
  \center{{\footnotesize {\bf Fig. \ref{fig:7}b.} Distribution of EAM at 3790-3810 \AA}}
  \end{minipage}
  \caption{The EAMs versus the number of stars. There are two peaks at $\sim$0.5\AA\, and $\sim$3.0\AA\, at 3700-3720 \AA. 
  On the other hand, there is only one peak at 0\AA\, for 3790-3810 \AA.}
  \label{fig:7}
  
  \begin{minipage}{0.5\hsize}
  \begin{center}
    \FigureFile(80mm,50mm){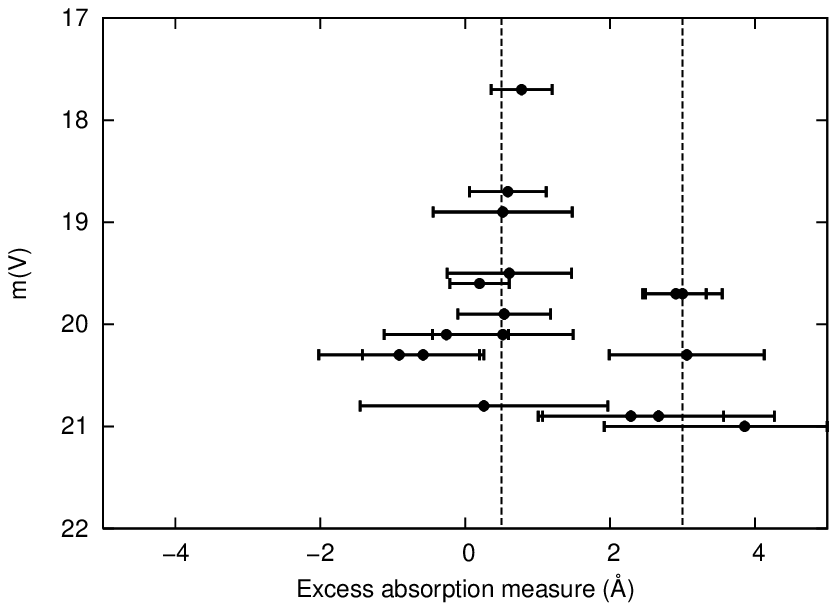}
  \end{center}
  \center{{\footnotesize {\bf Fig. \ref{fig:8}a.} Distribution of EAM at 3700-3720 \AA}}
  \end{minipage}
  \begin{minipage}{0.5\hsize}
  \begin{center}
    \FigureFile(80mm,80mm){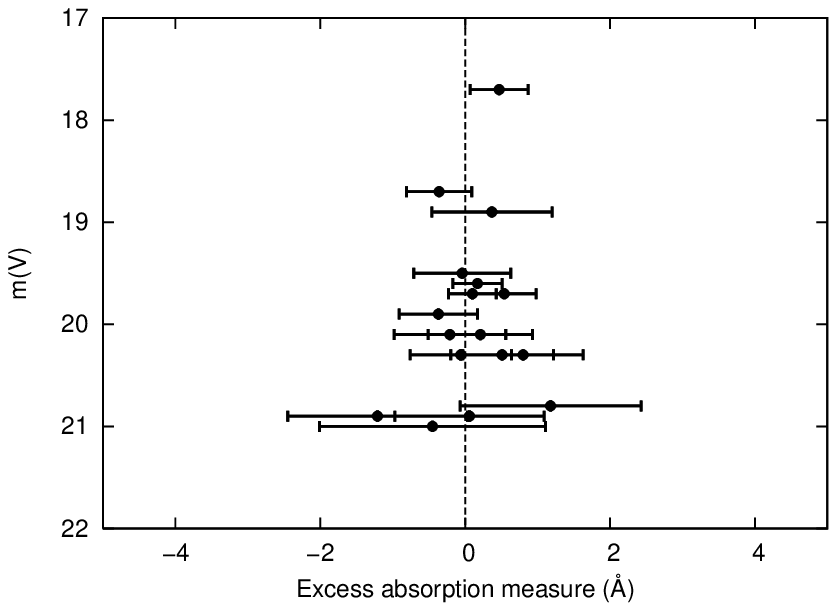}
  \end{center}
  \center{{\footnotesize {\bf Fig. \ref{fig:8}b.} Distribution of EAM at 3790-3810 \AA}}
  \end{minipage}
  \caption{The EAMs versus the magnitude in $V$-band. 
  The dashed vertical lines indicate the mean value given by Figure \ref{fig:7}. 
  The stars of about 21 mag have large errors.}
  \label{fig:8}
  
\end{figure*}

We present a statistical study of the broad absorption lines at around 3720 \AA\,. 
The EAM distribution of the observed stars is plotted in Figure \ref{fig:7} 
and the EAMs in the wavelength ranges of 3700-3720 \AA\, and 3790-3810 \AA\, are plotted 
with their magnitudes in $V$-band in Figure \ref{fig:8}. 

In 3700-3720 \AA\, a broad absorption line due to Fe I is expected 
not only in the spectrum of the companion star but also in the spectra of the background stars, 
while in 3790-3810 \AA\, no absorption is expected. 
In 3790-3810 \AA, the obtained EAM distribution is consistent with a scatter due to measurement errors 
around EAM $\sim$0 \AA\, (Figure \ref{fig:7}b and \ref{fig:8}b). 
On the other hand, there seem to be several stars which show positive EAMs of around 3 \AA\, 
in addition to stars without showing finite EAMs (Figure \ref{fig:7}a and \ref{fig:8}a). 

\begin{figure*}
  \begin{minipage}{0.5\hsize}
  \begin{center}
    \FigureFile(80mm,50mm){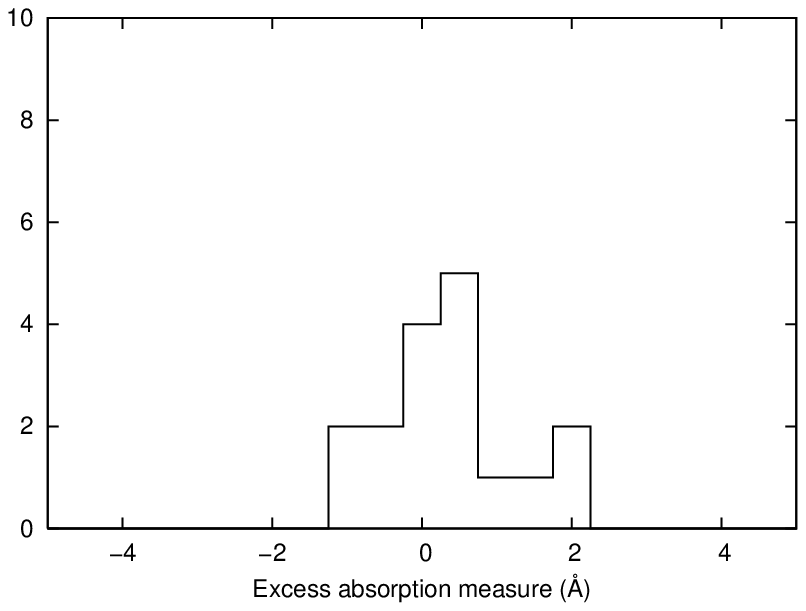}
  \end{center}
  \center{{\footnotesize {\bf Fig. \ref{fig:9}a.} Distribution of EAM at 3720-3740 \AA}}
  \end{minipage}
  \begin{minipage}{0.5\hsize}
  \begin{center}
    \FigureFile(80mm,80mm){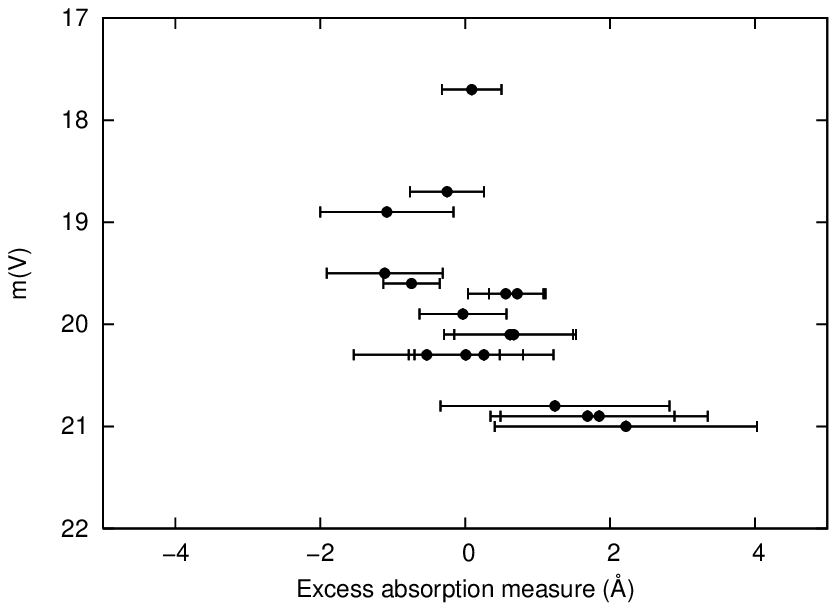}
  \end{center}
  \center{{\footnotesize {\bf Fig. \ref{fig:9}b.} EAM versus V-mag}}
  \end{minipage}
  \caption{Distribution of EAMs at 3720-3740 \AA\, same as Figure \ref{fig:7} and \ref{fig:8}. 
  We expect to detect the absorption lines from the background stars, 
  but we cannot detect them clearly like the EAMs at 3700-3720 \AA. 
  }
  \label{fig:9}
  
\end{figure*}

Next, we discuss the profiles of absorption lines. 
The profile in the spectrum of Tycho(E) was discussed in \S4. 
If Tycho(E) is the companion star, 
then the other stars with broad absorption lines should be located in the background of SNR Tycho.
These stars are expected to show broad absorption lines 
due to Fe I in the ejecta both in the blue and red sides of the original wavelength 3720 \AA. 
However, they exhibit very weak absorptions in the red side compared with that in the blue. 
In Figure \ref{fig:9}, we show the EAM distribution in the range 3720-3740 \AA\, similar to Figures \ref{fig:7} and \ref{fig:8}. 
There seem to be 4 stars having weak absorptions, but they are faint and the EAMs have large errors. 
In contrast, the other stars are distributed around 0 \AA. 

There are 5 stars outside of the candidate region from which broad absorption lines are detected. 
Tycho(ah) has the best signal-to-noise ratio and is already discussed in \S\ref{tychoh}. 
Tycho(aa) exhibits a similar spectrum but with smaller signal-to-noise ratios.
Tycho(ad), (ai), and (ak) seem to have broad absorption lines with larger EAMs in the red side of 3720 \AA\, than Tycho(ah). 
These EAMs of Tycho(ad), (ai), and (ak) are about a half of the EAMs of the corresponding blue side absorption lines. 
However, large errors involved in these EAMs present us from deducing a clear conclusion. 

If positive EAMs are not caused by spectral features of peculiar stars,
absorptions of Fe I is the most probable interpretation.
However, the observed absorptions are not exactly the same as predicted by the model of OS06. 
No red wings are significantly detected, 
while a significant blue wing is found for Tycho(ah) 
although its projected location is far from the center of SNR. 
Hence, if the origin of this absorption is Fe I, refining OS06 would be necessary. 
Their model assumes spherical symmetry distribution of the ejected iron. 
But if distribution of neutral iron is clumpy, 
the Fe I absorption might not be able to be detected according to a location of a star 
because of lack of the iron in the line of sight.
Also absorption profiles would be complicated.

Finally, we discuss Tycho(E) as candidates of the companion star of Tycho's SN.
Although there is no clear example which shows significant broad absorption lines 
both in blue and red sides of 3720 \AA\, which are expected for the background of the SNR, 
we confirm the blue-shifted broad absorption line at 3720 \AA\, only for Tycho(E) in the candidate region. 
An absorption line with this unique shape can be formed by Fe I in the SN ejecta when the star is inside SNR Tycho. 
If the supernova explosion occurred at the center of SNR Tycho defined by Warren et al. (2005), 
the projected distance of Tycho(E) is considerably short from the center of SNR Tycho ($\sim$10"), 
while Tycho(G) is located away from the center of SNR Tycho ($\sim$25''). 
Ruiz04 suggested that the SN site was about 25" off the geometrical center of the SNR. 
It is not clear whether this SN site is consistent with the observed shape of the SNR 
in the context of hydrodynamical evolution of the blast wave in the ambient inhomogeneous medium.
Therefore, it is likely that Tycho(E) is another candidate for the companion star, 
considering the position from the center of SNR Tycho. 

\section{Summary}
We have performed photometric and spectroscopic observations to  
search for the companion star remaining in the remnant of a SN Ia (SNR Tycho). 
First, we selected candidate stars for the spectroscopic observations from photometric observations 
for the SNR Tycho region. Then we obtained optical spectra of 4 stars in the candidate region 
and 13 other stars outside of the candidate region.
To identify the companion star of Tycho's SN, we compare the observed spectra in SNR Tycho with 
spectra of Ja84 at 3720 \AA, 3800 \AA, and 3860 \AA. 
If the position of a star is inside the ejecta of SNe Ia or in the background of the ejecta, 
absorption lines due to Fe I in the expanding ejecta may be seen. 

We are able to detect significant broad absorption lines at 3720 \AA\, for Tycho(E) and (ah). 
As the distribution of EAMs at 3720 \AA\, shows, spectra of Tycho(aa), (ad), (ai) and (ak) have possible absorption lines. 
Tycho(E) is located at the candidate region, while Tycho(aa), (ad) ,(ah), (ai) and (ak) are located out of the region.
However, we cannot detect the absorption at 3860 \AA\, at a sufficient significance level.  
It is difficult to detect them at 3860 \AA\, because of uncertainty of template stellar spectra at 3860 \AA. 

The detected absorption lines in the blue side at 3720 \AA\, are consistent with Fe I absorption of SNR Tycho.  
On the other hand, we do not obtain a clear broad absorption line in the red side of the original wavelength 
from a star in the background of the SNR. 
We cannot conclude if the detected broad absorption lines are due to Fe I in the SN ejecta. 
Another interpretation is that their absorption lines might be caused by the peculiarity of stars. 

We do not detect any broad absorption lines in the spectrum of Tycho(G). 
Thus we do not obtain supporting evidence for the result of Ruiz04 from our observations. 
In the candidate region, 
Tycho(E) is the only star within the candidate region that shows a significant broad absorption line.
The spectral type of Tycho(E) determined by our observations and by Ruiz04 are apparently inconsistent. 
Our observations indicate that Tycho(E) is a new candidate as a companion star  
in the sense that this star is located close to the projected center of the SNR and shows a broad absorption line.

\section*{ACKNOWLEDGMENTS}
We would like to thank the Subaru Telescope staff for their invaluable assistance. 
We thank M. Iye and N. Ebizuka for practical advice to use FOCAS in short wavelength. 
We are also grateful to W. Aoki for useful discussions about stellar profiles. 
We thank Dr. David Branch and an anonymous referee 
for constructive criticism which helped us to improve the paper. 
This work was supported in part with scientific research grants (15204012, 16540213, and 17104002) 
from the Ministry of Education, Science, Culture, and Sports of Japan. 


\end{document}